\documentclass[12pt]{article}
\usepackage[margin=1.in]{geometry}
\usepackage[utf8]{inputenc}
\usepackage{amsthm,amsmath,physics,mathtools,amssymb}

\usepackage{charter}

\usepackage{CJKutf8}

\usepackage{enumitem}

\usepackage{hyperref}
\hypersetup{
    colorlinks=true,
    linkcolor=blue,
    filecolor=magenta,
    urlcolor=cyan,
}
\usepackage[capitalise]{cleveref}

\theoremstyle{definition}

\usepackage{graphicx}
\graphicspath{{figure_arxiv/}}

\usepackage{tensor}

\DeclarePairedDelimiterX{\inp}[2]{\langle}{\rangle}{#1, #2}

\usepackage{xparse}

\NewDocumentCommand\LH{mo}{%
  \IfNoValueTF{#2}
   {\mathcal{L}(\mathcal{H}^{#1})}
   {\mathcal{L}(\mathcal{H}^{#1},\mathcal{H}^{#2})}%
}

\newcommand\id{\leavevmode\hbox{\small1\kern-3.3pt\normalsize1}}

\newcommand{\sV}{\mathbb{V}}
\newcommand{\sA}{\mathbb{A}}

\allowdisplaybreaks

\DeclareMathOperator\Log{Log}

\usepackage{bbold}

\usepackage{authblk}

\usepackage[title]{appendix}

\usepackage{mciteplus}

\usepackage{enumitem}

\title{Light ray fluctuations in simplicial quantum gravity}

\author{Ding Jia (贾丁)\thanks{djia@perimeterinstitute.ca}}
\affil{Perimeter Institute for Theoretical Physics, Waterloo, Ontario, N2L 2Y5, Canada}
\affil{Department of Physics and Astronomy, University of Waterloo, Waterloo, Ontario, N2L 3G1, Canada}
\date{}

\begin{document}

\begin{CJK*}{UTF8}{gbsn}
\maketitle
\end{CJK*}

\begin{abstract}
A non-perturbative study on the quantum fluctuations of light ray propagation through a quantum region of spacetime is long overdue. Within the theory of Lorentzian simplicial quantum gravity, we compute the probabilities for a test light ray to land at different locations after travelling through a symmetry-reduced box region in 2,3 and 4 spacetime dimensions. It is found that for fixed boundary conditions, light ray fluctuations are generically large when all coupling constants are relatively small in absolute value. For fixed coupling constants, as the boundary size is decreased light ray fluctuations first increase and then decrease in a 2D theory with the cosmological constant, Einstein-Hilbert and R-squared terms. While in 3D and 4D theories with the cosmological constant and Einstein-Hilbert terms, as the boundary size is decreased light ray fluctuations just increase. Incidentally, when studying 2D quantum gravity we show that the global time-space duality with the cosmological constant and Einstein-Hilbert terms noted previously also holds when arbitrary even powers of the Ricci scalar are added. We close by discussing how light ray fluctuations can be used in obtaining the continuum limit of non-perturbative Lorentzian quantum gravity.
\end{abstract}

\section{Introduction}

\begin{figure}
    \centering
    \includegraphics[width=.6\textwidth]{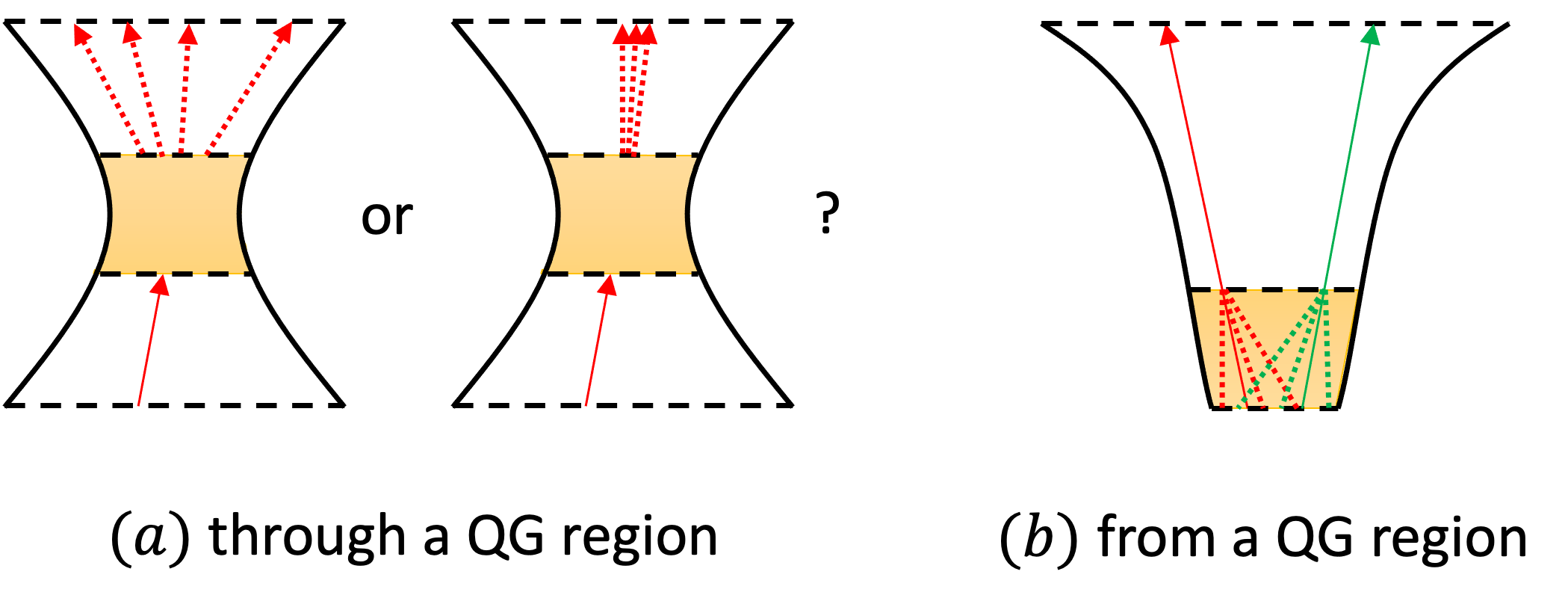}
    \caption{Light ray propagations affected by quantum gravitational (QG) regions.}
    \label{fig:lrfs}
\end{figure}

The topic of light ray fluctuations is relevant to some core themes of quantum gravity.

Consider the bounce scenario for quantum cosmology and quantum black holes \cite{Malafarina2017ClassicalReview} illustrated in \Cref{fig:lrfs} part (a). Suppose the cosmological and/or black hole singularities in classical gravity are replaced by bouncing processes in quantum gravity. A natural question is how light rays propagate through the shaded bouncing region where quantum gravitational effects are significant. After going through the quantum gravitational region, will a light ray become quantum dispersed as in the left picture, or stay focused as in the right picture? The answer influences not just our theoretical understanding of information propagation in black hole and cosmological physics, but also experimental searches of pre-bounce relics to test the bounce scenario.

As another example consider the propagation of light rays in cosmology illustrated in \Cref{fig:lrfs} part (b). Two light rays that were never in causal contact if spacetime was treated classically (they trace out the solid lines in the figure) could actually have been in causal contact if spacetime is treated quantumly to allow quantum fluctuations of the light ray paths (dashed lines). This difference could affect our qualitative and quantitative understanding on early universe cosmology \cite{Mukhanov2005PhysicalCosmology}, in particular on the horizon problem and the inflation hypothesis.

\begin{figure}
    \centering
    \includegraphics[width=.25\textwidth]{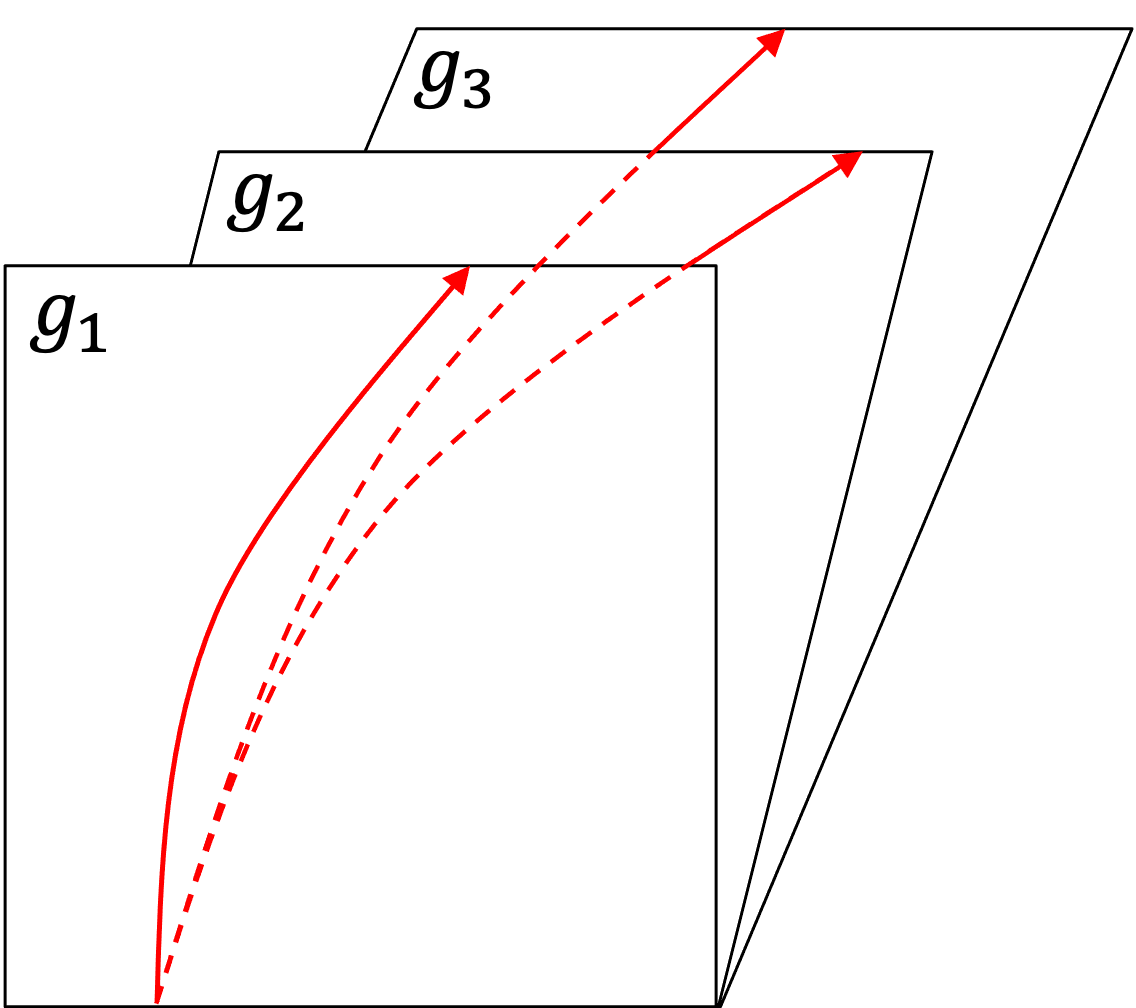}
    \caption{In a quantum region with superposed spacetime configurations, light rays starting at the same location on the boundary end at different locations on the other side of the boundary.}
    \label{fig:lrss}
\end{figure}

In general, we are interested in the propagation of light rays across of a region of quantum spacetime where different spacetime configurations are in superposition, yielding different paths of light ray propagation (\Cref{fig:lrss}). Since strong gravity is involved in the cosmology and black hole scenarios mentioned above, we are interested in a non-perturbative treatment of the problem.

Conceptually, it is very clear how to study the problem in non-perturbative Lorentzian gravitational path integrals. Let there be a region of quantum spacetime with fixed boundary configurations. Different spacetime configurations compatible with the boundary configuration are summed over, yielding amplitudes for the light ray to land at different locations.

Practically, how smoothly the study would proceed depends very much on which non-perturbative Lorentzian gravitational path integral is used. In spin-foam models, before proceeding it needs to be clarified whether the spin-foams represent continuum spacetime configurations or some fundamentally discrete structure. This choice affects where light rays and causal paths  \cite{BianchiCausalSpin-foams} can travel on a spin-foam (see \cite{DowkerRecoveringGravity} for a related discussion). 
In quantum causal set path integrals, it needs to be decided if the path integral should be restricted to configurations corresponding to a particular spacetime dimension, and if so how (see Section 6.4 of \cite{Surya2019TheGravity} for a discussion). In causal dynamical triangulation, the topic is more accessible since the light ray path on a piecewise flat spacetime configuration is obtainable and the path integral is well specified in different dimensions \cite{Ambjorn2012NonperturbativeGravity, Loll2020QuantumReview}. While we think it is possible to study the topic under discussion in causal dynamical triangulation, to our knowledge such studies have not been carried out before.

In this work we study light ray fluctuations across a quantum region of spacetime in Lorentzian simplicial quantum gravity \cite{Jia2022ComplexProspects}. The study of gravitational path integrals defined in terms of simplicial spacetime configurations {\'a} la Regge \cite{Regge1961GeneralCoordinatesb} has a long history \cite{Rocek1981QuantumCalculus, Williams1992ReggeBibliography, Loll1998DiscreteDimensions, Hamber2009QuantumApproach, Barrett2019TullioGravity}. While previous works focused on the Euclidean theory, there has been a growing interest in the Lorentzian theory in recent years \cite{Tate2011Fixed-topologyDomain, Tate2012Realizability1-simplex, Asante2021EffectiveGravity, Dittrich2022LorentzianSimplicial, Jia2022Time-spaceGravity, Jia2022ComplexProspects, AsanteComplexCosmology}. As in causal dynamical triangulation, the light ray paths on a piecewise flat simplicial spacetime configuration is obtainable, in particular using Lorentzian trigonometry which we illustrate in \Cref{sec:lrl}. From there one could integrate over spacetime configurations corresponding to different light ray landing locations to obtain the quantum amplitudes and the probabilities.

The non-perturbative Lorentzian path integral is not easy to compute due to the complex phase of the integrand. To facilitate the study we make two simplifications. First, we consider the fluctuation of \textit{test} light rays. In other words we consider models of pure quantum gravity to infer the light paths from the gravitational configurations alone, without introducing matter degrees of freedom. This means the backreaction of light on gravity is not taken into account in this simplified study. Second, we focus on a symmetry-reduced ``box model'' with simple boundary conditions and only one dynamical degree of freedom. This allows us to evaluate the Lorentzian path integral through direct numerical integration.

In this simplified model, we ask how light ray fluctuations are affected by the coupling constants and the size of the region determined by the boundary conditions in 2,3 and 4 spacetime dimensions. For fixed boundary conditions, we find that light ray fluctuations are generically large when all coupling constants are relatively small in absolute value in all dimensions. For fixed coupling constants, we find that as the boundary size is decreased, light ray fluctuations first increase and then decrease in a 2D theory with the cosmological constant, Einstein-Hilbert and $R^2$ terms. On the other hand, as the boundary size is decreased light ray fluctuations just increase in 3D and 4D theories with the cosmological constant and Einstein-Hilbert terms.
As a side result, when studying 2D quantum gravity we show that the global time-space duality with the cosmological constant and Einstein-Hilbert terms noted previously \cite{Jia2022Time-spaceGravity} also applies when arbitrary even powers of the Ricci scalar are added.

These results point to light ray fluctuations as potentially useful in the study of the renormalization group and the continuum limit of non-perturbative Lorentzian quantum gravity. In performing renormalization group type analysis by refining the lattices to approach the continuum limit, it is important to find physical quantities to compare across different lattices. The present study reveals the light ray amplitudes and light ray probabilities as candidate physical quantities to compare across lattices. These quantities which are not accessible in the Euclidean offer some new opportunities to be explored in future works.

The paper is organized as follows. The formalism of Lorentzian simplicial quantum gravity is introduced in \Cref{sec:lsqg}. The symmetry-reduced box model and the formulas for light ray locations across the box region are presented in \Cref{sec:srbm}. The results in 2,3 and 4D are presented in \Cref{sec:lrf2d} to \Cref{sec:lrf4d}. The important quantity of the light ray amplitude and its relevance to the continuum limit of Lorentzian quantum gravity is discussed in \Cref{eq:lacl}. A brief summary including discussions on future prospects is given in \Cref{sec:d}.

\section{Lorentzian simplicial quantum gravity}\label{sec:lsqg}

\begin{figure}
    \centering
    \includegraphics[width=.2\textwidth]{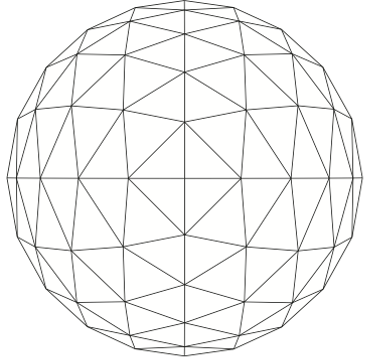}
    \caption{Describing curved space/spacetime by gluing flat simplicies.}
    \label{fig:lcf_simplicial-geometry}
\end{figure}

Formally, gravitational path integrals take the form
\begin{align}\label{eq:qgl}
Z=&\int \mathcal{D}g ~ A[g]
\end{align}
of a sum over gravitational configurations $g$ weighted by amplitudes $A[g]$. To fully define the path integral, we need specify a way to enumerate the gravitational configurations to perform the sum. In simplicial quantum gravity \cite{Regge1961GeneralCoordinatesb, Rocek1981QuantumCalculus, Williams1992ReggeBibliography, Loll1998DiscreteDimensions, Hamber2009QuantumApproach, Barrett2019TullioGravity}, the sum is over simplicial spacetime configuration which describe curved spacetime by combining flat simplicies (\Cref{fig:lcf_simplicial-geometry}). While extensive works have been carried out in the past in the Euclidean signature, the present study of light ray fluctuations is based on a Lorentzian version of the theory \cite{Jia2022ComplexProspects} that sums Lorentzian simplicial spacetime configurations \cite{Sorkin1975Time-evolutionCalculus, SorkinLorentzianVectors}.


In classical General Relativity a spacetime configuration is characterized by the metric field $g_{ab}$. Its physical meaning is that the line element $ds^2 = g_{ab} dx^a dx^b$ indicates the squared length between infinitesimally separated points. In simplicial gravity a spacetime configuration is characterized by the \textbf{squared lengths}
\begin{align}\label{eq:slfg}
\sigma_e=\int_e ds^2
\end{align}
integrated along the simplicial edges $e$. As such, $\sigma_e$ is the finite version of the line element $ds^2$. In the metric signature
\begin{align}
(-,+,\cdots,+)
\end{align}
used here, $\sigma_e$ can be smaller than, equal to, or greater than zero, corresponding to the edge being timelike, lightlike, and spacelike. The spacetime within a simplex is taken to be flat, and the simplicial configuration is fully characterized by $\sigma$ on all the edges.

The gravitational path integral is then a sum over simplicial spacetime configurations specified by the edge squared lengths on simplicial lattice graphs $\Gamma$:
\begin{align}\label{eq:pf}
Z =& \int \mathcal{D}\sigma ~ e^{E[\sigma]},
\\
\int \mathcal{D}\sigma =&  \lim_\Gamma\prod_{e\in\Gamma} \int_{-\infty}^\infty d\sigma_e ~ \mu[\sigma] L[\sigma] C[\sigma].\label{eq:sqgm1}
\end{align}
On a finite lattice $\Gamma$ the integral gives an approximate result. The exact result is approached in the limit $\lim_\Gamma$ of infinitely refining the lattice.\footnote{It is possible to include an additional sum over spacetime topologies by summing over lattices with different topologies in \eqref{eq:sqgm1}. Yet in the simple box models studied below we will focus on spacetime configurations with the trivial topology.}
The symbols $L[\sigma], C[\sigma], \mu[\sigma], E[\sigma]$ stand for the Lorentzian constraint, the lightcone constraint, the integration measure factor, and the path integral exponent. Their forms are specified below.

\subsection{Path integral exponent}\label{sec:pie}

The formal continuum path integral exponent takes the form
\begin{align}\label{eq:E2D}
E =& i \int d^D x \sqrt{-g} ( - \lambda +k R + a R^2 +\cdots),
\end{align}
where $\lambda, k, a$ are coupling constants and $\cdots$ signifies the possibility of including additional terms. We want to find the simplicial versions of the exponent.

\subsection*{2D}\label{sec:2D}

In 2 spacetime dimensions the simplicial path integral exponent takes the form
\begin{align}
    E = & - \lambda \sum_t A_t - k \sum_v \delta_v + a \sum_v \frac{\delta_v^2}{A_v} +\cdots.
\end{align}
The first term is the cosmological constant term. For a triangle $t$ with squared edge lengths $\sigma _{01},\sigma _{02},\sigma _{12}$, the squared area formula
\begin{align}\label{eq:2dsvol}
\sA_t=&\frac{1}{16} \left(-\sigma _{01}^2-\sigma _{02}^2-\sigma _{12}^2+2  \sigma _{01} \sigma _{02}+2  \sigma _{01} \sigma _{12}+2 \sigma _{02} \sigma _{12}\right)
\end{align}
generalizes Heron's formula to apply to both Euclidean and Lorentzian cases. $\sA_t$ is positive in the Euclidean and negative in the Lorentzian, just like the squared area factor $g=\det g_{ab}$. The Lorentzian triangle area
\begin{align}\label{eq:2dvol}
A_t =& \sqrt{\sA_t}
\end{align}
is positive imaginary and forms the analogue of $\sqrt{g}=i\sqrt{-g}$. This explains why $- \lambda \sum_t A_t$ is the simplicial version of $-i \lambda \int d^D x \sqrt{-g}$.

The second term is the Einstein-Hilbert term. Simplicial gravity is based on the idea that composing flat simplicies can describe curved spacetime configurations. This is possible because the sum of angles around a vertex can differ from the flat spacetime value. The difference is encoded in the \textbf{deficit angle}
\begin{align}\label{eq:da1}
\delta_v =& F_v - \sum_{t\ni v}\theta_{t, v}.
\end{align}
In this formula, $\sum_{t\ni v}\theta_{t, v}$ is the sum of triangle angles $\theta_{t, v}$ around a vertex $v$, while $F_v$ is flat spacetime value. If $v$ lies in the interior of a region, then as explained below $F_v=2\pi$. If $v$ lies on the boundary of a region, then $F_v$ depends on how many pieces of spacetime regions share the vertex $v$, and can only be fixed on a case by case basis. To obtain a lattice version of the Ricci scalar $R$ we even out $\delta_v$ to unit areas through $-\delta_v/A_v$ where \begin{align}\label{eq:av}
A_v =& \frac{1}{3}\sum_{t\ni v} A_t.
\end{align}
is the vertex share of the Lorentzian area. A triangle $t$ contains three vertices $v$, so a vertex shares $1/3$ of the triangle's area $A_t$. In the continuum limit $-\delta_v/A_v$ approaches $R$ for $\theta$ defined in \eqref{eq:theta} \cite{Sorkin1975Time-evolutionCalculus, Jia2022ComplexProspects}.\footnote{Here the sign convention for the curvature is set by ${R^{\rho }}_{\sigma \mu \nu }=\partial _{\mu }\Gamma _{\nu \sigma }^{\rho }-\partial _{\nu }\Gamma _{\mu \sigma }^{\rho }+\Gamma _{\mu \lambda }^{\rho }\Gamma _{\nu \sigma }^{\lambda }-\Gamma _{\nu \lambda }^{\rho }\Gamma _{\mu \sigma }^{\lambda }$.} This explains why $-k \sum_v A_v \frac{\delta_v}{A_v}=-k \delta_v$ forms the lattice version of the Einstein-Hilbert term $i k \int d^D x \sqrt{-g} R$.
In 2D a Lorentzian version of the Gauss-Bonnet theorem \cite{SorkinLorentzianVectors} indicates that the Einstein-Hilbert term is a topological invariant. Therefore it can usually be taken out of the path integration to simplify the study.

The third term $a \sum_v \frac{\delta_v^2}{A_v}=a A_v (-\sum_v \frac{\delta_v}{A_v})^2$ is the lattice version of $a \int d^D x ~i\sqrt{-g} R^2$, given that $-\delta_v/A_v$ is the lattice analogue of $R$. This term is not a topological invariant and allows non-trivial stationary points for the action. In this work, we focus on these first three terms in the exponent for the studies.

\begin{figure}
    \centering
    \includegraphics[width=.3\textwidth]{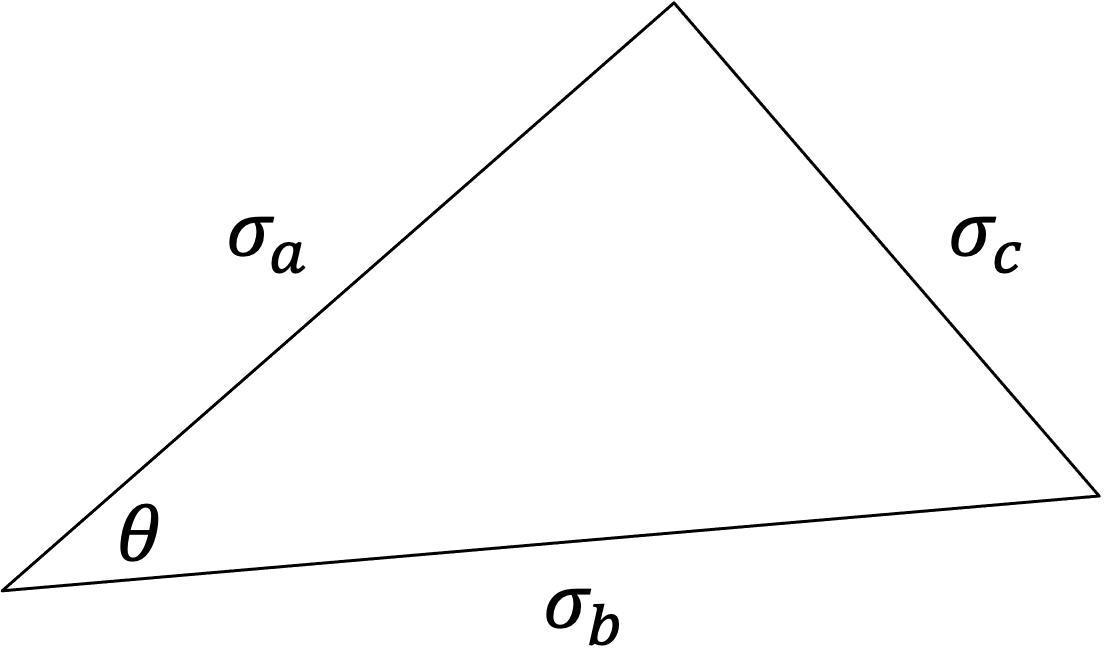}
    \caption{Angle of a Lorentzian triangle.}
    \label{fig:lcf_triangle}
\end{figure}

Since the integration variable is the squared length, we need to express the deficit angle $\delta_v$ and hence the triangle angles $\theta_{t, v}$ in squared length. For a Lorentzian or Euclidean triangle with squared lengths as shown in \Cref{fig:lcf_triangle} \cite{Jia2022ComplexProspects},
\begin{align}
\theta =& -i \Log \alpha,\label{eq:theta}
\\
\alpha=& \frac{\sigma_{a}+\sigma_{b}-\sigma_c+\sqrt{\sigma _{a}^2+\sigma _{b}^2+\sigma _{c}^2-2  \sigma _{a} \sigma _{b}-2  \sigma _{b} \sigma _{c}-2 \sigma _{c} \sigma _{a}}}{-2\sqrt{-\sigma_a}\sqrt{-\sigma_b}},\label{eq:alpha}
\end{align}
where $\Log$ stands for the principle branch of the log function ($\Im\Log\alpha=\pi$ for $\alpha<0$). To understand this formula, consider for a moment an Euclidean triangle with squared lengths $\sigma_{a},\sigma_{b},\sigma_c$. In this case the first part of $\alpha$, $\frac{\sigma_{a}+\sigma_{b}-\sigma_c}{-2\sqrt{-\sigma_a}\sqrt{-\sigma_b}}$, is simply $\cos\theta$ according to the law of cosines. The second part, $\frac{\sqrt{\sigma _{a}^2+\sigma _{b}^2+\sigma _{c}^2-2  \sigma _{a} \sigma _{b}-2  \sigma _{b} \sigma _{c}-2 \sigma _{c} \sigma _{a}}}{-2\sqrt{-\sigma_a}\sqrt{-\sigma_b}}$, is simply $i\sin\theta$ by recognising the numerator as $4i$ times the triangle area according to Heron's formula. Therefore $\theta = -i \Log \alpha$ holds for an Euclidean triangle. By allowing the squared lengths to be negative we arrive at the angle formula \eqref{eq:theta} which applies in both the Lorentzian and the Euclidean.

It can be check that in flat spacetime the angles around a vertex sum to $2\pi$ \cite{Jia2022ComplexProspects}, which confirms the claim above about $F_v$.

\subsection*{Higher dimensions}

In higher dimensions, we consider the simplicial path integral exponent
\begin{align}\label{eq:EgD}
E = & - \lambda \sum_s \sqrt{\sV_s} + i k \sum_h \delta_h \sqrt{-\sV_h} + \cdots.
\end{align}
The first term is the cosmological constant term. For a $D$-simplex $s$ with squared edge lengths $\sigma _{01},\sigma _{02},\cdots$, the Cayley-Menger determinant
\begin{align}\label{eq:svol}
\sV_s = \frac{(-1)^{D+1}}{2^D (D!)^2}
\begin{vmatrix}
 0 & 1 & 1 & 1 & \ldots  & 1 \\
 1 & 0 & \sigma _{01} & \sigma _{02} & \ldots  & \sigma _{0 d} \\
 1 & \sigma _{01} & 0 & \sigma _{12} & \ldots  & \sigma _{1 d} \\
 1 & \sigma _{02} & \sigma _{12} & 0 & \ldots  & \sigma _{2 d} \\
 \vdots  & \vdots  & \vdots  & \vdots  & \ddots & \vdots  \\
 1 & \sigma _{0 d} & \sigma _{1 d} & \sigma _{2 d} & \ldots  & 0 \\
\end{vmatrix}.
\end{align}
yields its squared volume, which generalizes \eqref{eq:2dsvol} to $D$ dimensions. $\sV_s$ is the analogue of $g$ and both are positive in the Euclidean and negative in the Lorentzian. This explains why $- \lambda \sum_s \sqrt{\sV_s}$ is the simplicial version of $-i \lambda \int d^D x \sqrt{-g}$.

The second term is the Einstein-Hilbert term. Imagine we generate a D-dimensional simplicial configurations by extending a 2D Lorentzian  configuration uniformly in D-2 additional spatial dimensions. The Einstein-Hilbert term would be $-k \sum_v\delta_v V$, which is the 2D term $-k\sum_v \delta_v$ times $V>0$, the spatial volume extended in the D-2 additional dimensions.
Given a general D-dimensional Lorentzian simplicial configuration, we could imagine that this is arrived at by non-uniformly extending a 2D configuration. The Einstein-Hilbert term would be
\begin{align}
i k \sum_h \delta_h \sqrt{-\sV_h}.
\end{align}
Here the sum is over codimension 2 subsimplices (edges in 3D, triangles in 4D etc.) referred to as \textbf{hinges} and labelled by $h$. At each hinge the deficit angle $\delta_h$ is obtained by projecting the $D$-simplices containing the hinge to the 2D plane orthogonal to the hinge and computing the deficit angle at the vertex where the hinge projects to in this plane (see \cite{Jia2022ComplexProspects} for the formula of the deficit angle in terms of squared lengths). The plane can be Euclidean or Lorentzian, and \eqref{eq:da1}, \eqref{eq:theta}, \eqref{eq:alpha} apply equally well given the squared lengths. As suggested above by the multiplication by $V$, the hinges are extended non-uniformly in D-2 dimensions so we multiply hingewise by the volumes $-i\sqrt{-\sV_h}$, where $\sV_h$ is the squared volume defined by \eqref{eq:svol} which applies to both Euclidean and Lorentzian hinges. This explains $i k \sum_h \delta_h \sqrt{-\sV_h}$ as the Einstein-Hilbert term for the extended configuration, which has the correct continuum limit \cite{Sorkin1975Time-evolutionCalculus, Jia2022ComplexProspects}.

In contrast to 2D, in higher dimensions the Einstein-Hilbert term is not a topological invariant, so we will ignore higher order terms in the following given that the first two terms already yields a non-trivial theory.

\subsection{Measure factor}

For the integration measure factor $\mu[\sigma]$, a commonly used family of local measures is the product of powers of the simplicial (square) volumes
\begin{align}\label{eq:mf}
\mu[\sigma]=\prod_s \sV_s^{m}
\end{align}
parametrized by $m\in\mathbb{R}$ \cite{Hamber2009QuantumApproach}.\footnote{Since $\sV_s$ is negative in the Lorentzian, one could use $\mu[\sigma]=\prod_s (-\sV_s)^m$ instead to make the measure positive. However, this makes no essential difference on a fixed lattice because the two measures differ at most by an overall constant.} In this work we adopt \eqref{eq:mf} as the measure factor. The constant $m$, like the coupling constants $\lambda,k$ and $a$, is treated as a parameter of the theory.

\subsection{Lorentzian and lightcone constraints}\label{sec:llc}


Without the Lorentzian constraint $L[\sigma]$, the integral \eqref{eq:sqgm1} also includes non-Lorentzian configurations such as Euclidean ones where all edges have positive square lengths and all simplices have positive squared volumes. The Lorentzian constraint $L[\sigma]$ serves to ensure that only Lorentzian spacetime configurations are included in \eqref{eq:sqgm1}.


A simplex $s$ is Lorentzian, i.e., embedable in 2D Minkowski spacetime, if and only if \cite{Tate2012Realizability1-simplex, Asante2021EffectiveGravity}
\begin{align}\label{eq:lgti}
\sV_s < 0; \text{ and } \sV_{r}<0 \implies \sV_{t}\le 0 \text{ for all $t\supset r$}.
\end{align}
This says that: 1) The simplex $s$ itself has negative squared volume, and 2) if a subsimplex $r$ of $s$ has negative squared volume, then all higher-dimensional subsimplices $t$ that contain $r$ have non-negative squared volumes.

The first condition is easy to digest since we know that the metric determinant $g$ as the infinitesimal squared volume is negative in the Lorentzian and positive in the Euclidean. The second condition is there because a timelike subsimplex cannot be embedded in a higher dimensional spacelike subsimplex (e.g., for a 4-simplex $s$, a timelike triangle subsimplex $r$ cannot be embedded in a spacelike tetrahedron subsimplex $t$), and the first condition is not enough to ensure this.
The Lorentzian constraint is then
\begin{align}\label{eq:lcd}
L[\sigma]=
\begin{cases}
1, \quad \text{if \eqref{eq:lgti} holds for all simplices }s,
\\
0, \quad \text{otherwise}.
\end{cases}
\end{align}

\begin{figure}
    \centering
    \includegraphics[width=.2\textwidth]{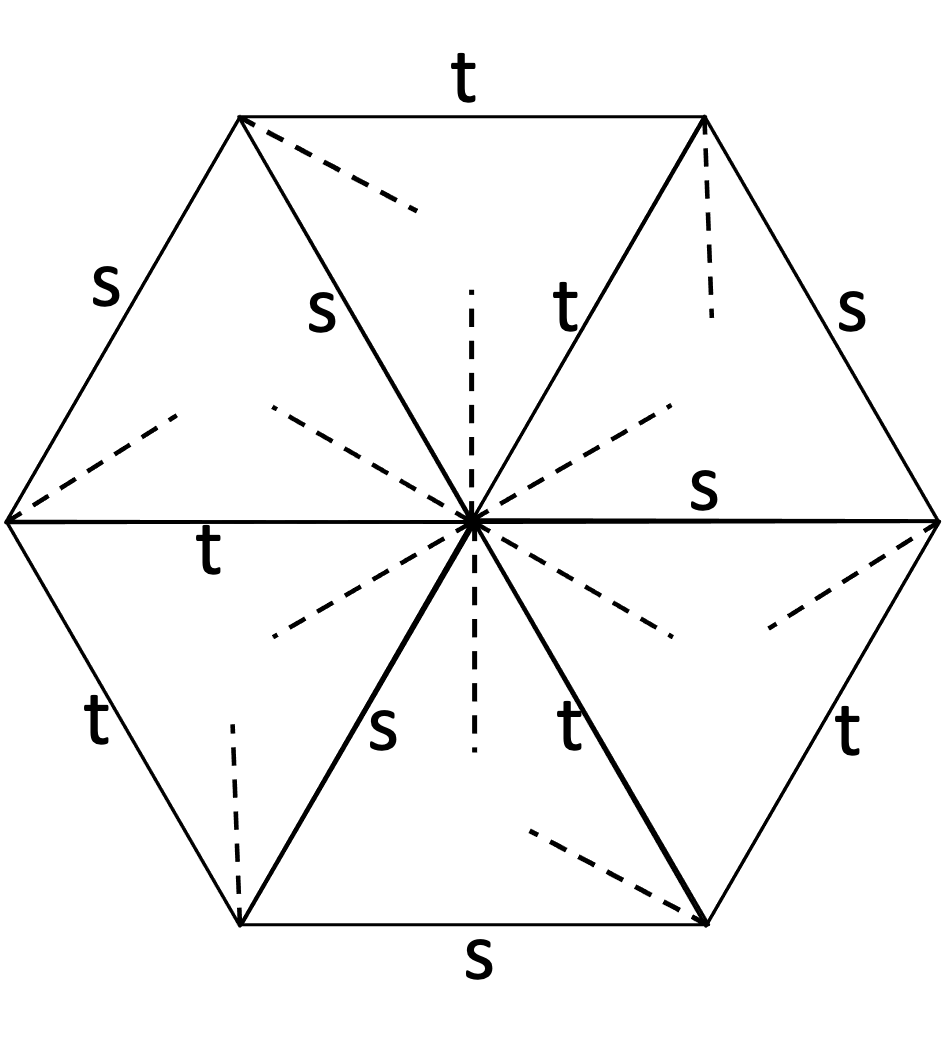}
    \caption{With s representing spacelike edges and t representing timelike edges, the vertex at the center of the figure has six light rays (dashed lines) and three lightcones.}
    \label{fig:2dlcs}
\end{figure}

The Lorentzian constraint is not enough to ensure the path integral includes only ordinary spacetime configurations. In an ordinary spacetime configuration, each point has two lightcones attached to it. Without the lightcone constraint $C[\sigma]$, the integral \eqref{eq:sqgm1} can include configurations where a vertex can have fewer or more than two lightcones, with the latter case illustrated in the 2D configuration of \cref{fig:2dlcs}. The lightcone constraint
\begin{align}
C[\sigma]=
\begin{cases}
1, \quad \text{all interior points have two lightcones attached},
\\
0, \quad \text{otherwise}
\end{cases}
\end{align}
ensures that all points in the interior region of the path integral configurations have two lightcones attached.

Since each simplex is just a portion of Minkowski spacetime, the lightcone number is always 2 in the interior of the simplices. On a boundary point of a simplex where multiple simplices meet, we need to count lightcones for each simplex and add up the number to check the lightcone constraint. For instance the center vertex in the 2D configuration of \cref{fig:2dlcs} is met by six triangles each coming with a light ray. Hence there are 3 lightcones, violating the lightcone constraint.

\subsection{Scaling identity}\label{sec:si}

Later we will study the dependence of the path integral on the boundary edge squared lengths. For this purpose it is useful to derive a scaling identity that relates a scaling of the boundary condition to a scaling of the coupling constants. Consider the path integral on a fixed lattice graph\footnote{If one considers lattice refinement to take the continuum limit, then the scaling identity \eqref{eq:si} may receive modifications for anomalous scaling dimensions. Although a brief schematic discussion on the continuum limit is given in \Cref{eq:lacl}, the computations performed in this work are based on fixed lattices so we will not consider anomalous dimensions in the scaling identity.} $\Gamma$ with fixed boundary squared lengths $\sigma_B$
\begin{align}
Z[\sigma_B,c_i,m] =& \prod_{e\in\Gamma} \int_{\sigma_B} d\sigma_e ~ L[\sigma] C[\sigma] ~ (\prod_s \sV_s^{m}) e^{\sum_i c_i E_i},\label{eq:exponent}
\end{align}
where the exponent is expressed with the coupling constants $c_i$ of length dimension $d_i$. Rewriting $\sigma$ on all edges as $l^2\sigma'$ where $l$ is an arbitrary constant number yields
\begin{align}
Z[\sigma_B,c_i,m] =& \prod_{e\in\Gamma} \int_{l^2\sigma_B'} d(l^2\sigma'_e) ~ L[l^2\sigma'] C[l^2\sigma'] ~ (\prod_s (l^{2D} \sV_s')^{m}) e^{\sum_i c_i l^{-d_i} E_i'}
\\
=& l^{2N_e+2m D N_s} \prod_{e\in\Gamma} \int_{\sigma_B'} d\sigma'_e ~ L' C' ~ (\prod_s (\sV_s')^{m}) e^{\sum_i c_i l^{-d_i} E_i'}
\\
=&l^{2N_e+2m D N_s} ~ Z[l^{-2}\sigma_B,c_i l^{-d_i},m].\label{eq:si}
\end{align}
Here $D$ is the spacetime dimension, $N_e$ is the number of non-boundary edges in $\Gamma$, $N_s$ is the number of simplices, and $f'$ is the shorthand for $f[\sigma']$ for a function $f[\sigma]$. In the second line we noted that the constraints $L$ and $C$ take the same values for $\sigma$ and $\sigma'$.

Setting $l^2 \sigma_B$ in place of $\sigma_B$ for \eqref{eq:si}, we get that
\begin{align}
Z[l^2 \sigma_B,c_i,m] =& l^{2N_e+2m D N_s} ~ Z[\sigma_B,c_i l^{-d_i},m].\label{eq:si2}
\end{align}
This says that changing the boundary condition by an arbitrary factor $l^2$ (LHS) can equivalently be achieved by scaling the coupling constants while keeping the boundary condition fixed (RHS).

\section{Symmetry-reduced box model}\label{sec:srbm}


\begin{figure}
    \centering
    \includegraphics[width=.8\textwidth]{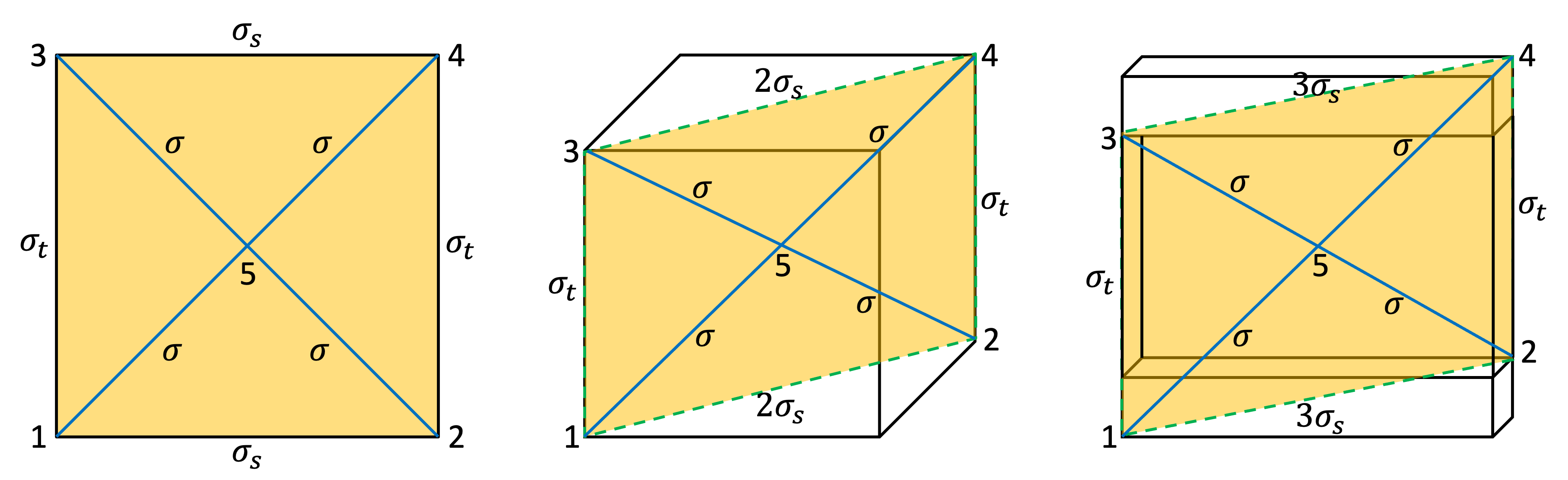}
    \caption{Symmetry-reduced box models in 2,3 and 4 spacetime dimensions. The diagonal 2D plane where the light ray travels is shaded and its boundary and interior squared lengths are labelled.}
    \label{fig:lcf_symmetry-reduced}
\end{figure}

A box model is a hypercube with an interior vertex connected to each of the boundary vertices by an edge. The hypercube is thus divided into flat hyperpyramids formed by the interior vertex as the tip and a face of the hypercube as the base. Here we focus on symmetry-reduced box models in 2,3 and 4 spacetime dimensions illustrated in \Cref{fig:lcf_symmetry-reduced}. The boundary condition is that:
\begin{align}
\text{All boundary spacelike edges have squared length $\sigma_s>0$;}\nonumber
\\
\text{All boundary timelike edges have squared length $\sigma_t<0$.}
\end{align}
In a path integral configuration all the interior squared lengths take the same value $\sigma$ and in this sense the model is symmetry-reduced.

In computing the path integral exponent, we assume the box to be a standalone region without neighbors. Therefore in 2D the vertex area $A_v$ of \eqref{eq:av} is the sum of two triangles. 
In addition, $F_v$ in the definition of the deficit angle \eqref{eq:da1} is fixed to take the value for a flat hypercube in general dimensions. For example in 2D $F_v=\pi/2$ as a quarter of the full angle $2\pi$.

Given that the hyperpyramids are not simplices, they need to be divided into simplices before applying the formulas of \Cref{sec:pie}. Alternatively the flat hyperpyramids could be treated as the elementary building blocks that describe curved spacetime configurations by having non-vanishing deficit angles where they are glued together. Since the hyperpyramids are flat, in the simplex description their interior deficit angles vanish. Hence the simplex description agrees with the pyramid description on the Einstein-Hilbert term. Since non-squared Lorentzian and Euclidean volumes are both additive, the two descriptions also agree on the cosmological constant term. The only difference lies in the measure term, which in the simplex description is given by \Cref{eq:mf} as $\mu[\sigma]=\prod_s \sV_s^{m}$. In the pyramid description this is replaced by
\begin{align}
\mu[\sigma]=\prod_p \sV_p^{m},
\end{align}
where the sum is over hyperpyramids $p$, and the hyperpyramid squared volume $\sV_p$ is negative for Lorentzian hyperpyramids. $\sV_p$ equals $-(\sum_{s\in p} \sqrt{-\sV_s})^2$ in terms of the simplex decomposition of the hyperpyramid. Below we conform with the pyramid description for the integration measure term, because the integrated variables are the squared edges on the boundaries of the hyperpyamid instead of the simplices.



\subsection{Light ray locations}\label{sec:lrl}

\begin{figure}
    \centering
    \includegraphics[width=.65\textwidth]{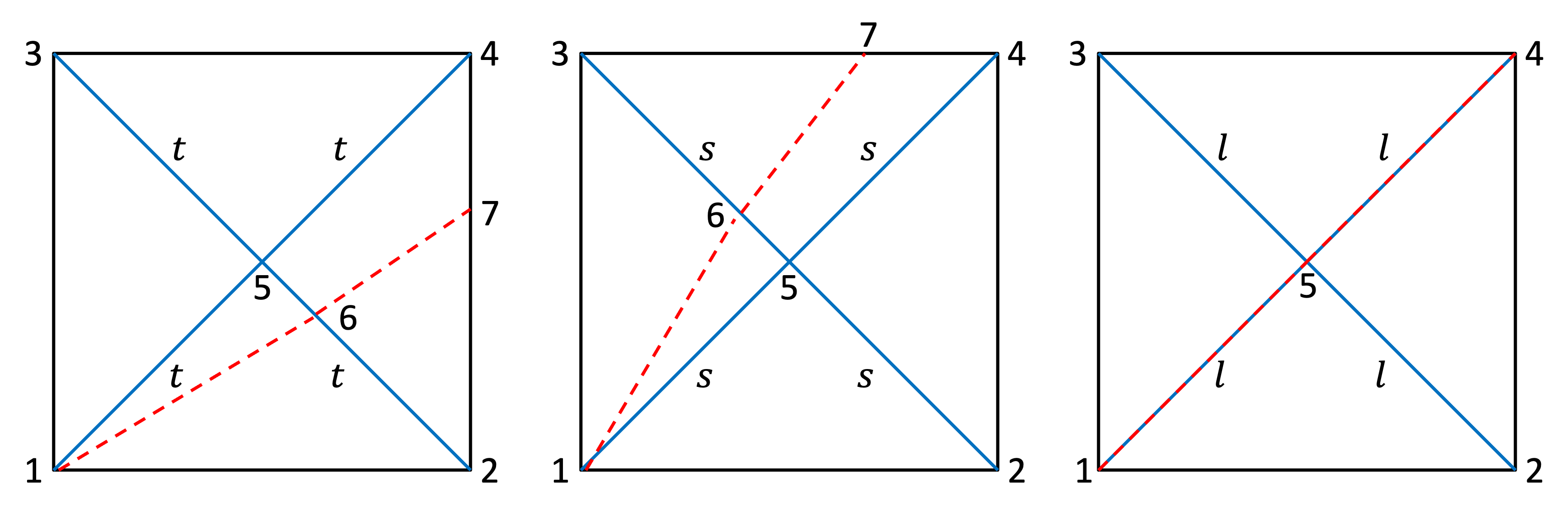}
    \caption{Situations for the light ray (dashed lines) emanating from vertex 1 in the symmetry-reduced box model when the interior edge is timelike ($t$), spacelike ($s$), and lightlike ($l$).}
    \label{fig:lcf_symmetry-reduced-box_light-ray}
\end{figure}

In all dimensions, we consider a light ray travelling through the quantum spacetime region within the diagonal 2D planes shaded in  \Cref{fig:lcf_symmetry-reduced}, again starting from vertex 1. In 2D, the spacelike edges of the plane have squared length $\sigma_s$. In 3D, the spacelike edges of the plane have squared length $2\sigma_s$ because they are the diagonal edges of the 2D base squares. In 4D, the spacelike edges of the plane have squared length $3\sigma_s$ because they are the diagonal edges of the 3D base cubes.

Therefore in all dimensions, the light ray travels through a 2D simplicial configuration consisting 4 triangles as illustrated in \Cref{fig:lcf_symmetry-reduced-box_light-ray}. Depending on whether the interior edge is timelike or spacelike, the light ray will land on either edge 24 or edge 34. If the interior edge is lightlike, the light ray will land right on the vertex 4.

To find the precise light ray landing location we apply Lorentzian trigonometry to triangles with one lightlike edge. Consider a triangle with squared lengths $\sigma_a$, $\sigma_b$ and $\sigma_c$ where edge $c$ is lightlike. For the angle bounded by the edges $a$ and $b$, setting $\sigma_c=0$ in \eqref{eq:alpha} yields
\begin{align}\label{eq:light-alpha}
\alpha^2=&(\frac{\sigma_{a}+\sigma_{b}+\sqrt{\sigma _{a}^2+\sigma _{b}^2-2  \sigma _{a} \sigma _{b}}}{-2\sqrt{-\sigma_a}\sqrt{-\sigma_b}})^2=(\frac{\sigma_{a}+\sigma_{b}+\abs{\sigma_a-\sigma_b}}{-2\sqrt{-\sigma_a}\sqrt{-\sigma_b}})^2=\frac{\sigma_+}{\sigma_-},
\end{align}
where $\sigma_+$ is the larger real number between $\sigma_a$ and $\sigma_b$, while $\sigma_-$ is the smaller one ($\sigma_a=\sigma_b$ is impossible because $\sA_t$ would be zero which violates the Lorentzian constraint of \Cref{sec:llc}).

For a box with timelike interior edges illustrated in the left of \Cref{fig:lcf_symmetry-reduced-box_light-ray}, $\sigma_{27}<\sigma_{26}<0<\sigma_{12}$. The first inequality holds because the future-pointing light ray 67 is moving ``upwards'' such that edge 27 has a larger timelike length than edge 26, which implies $\sigma_{27}<\sigma_{26}$ for the negatively signed squared lengths. Applying \eqref{eq:light-alpha} to triangles 126 and 267 yields $\sigma_{26}=\sigma_{12}/\alpha_{126}^2=\sigma_{12}/\alpha_{125}^2$ and $\sigma_{27}=\sigma_{26}/\alpha_{627}^2=\sigma_{26}/\alpha_{425}^2$, where
\begin{align}
\text{$\alpha_{ijk}$ refers to $\alpha$ of \eqref{eq:alpha} for the angle bounded by edges $ij$ and $kj$.}
\end{align}
Both $\alpha_{125}$ and $\alpha_{425}$ can be determined using \eqref{eq:alpha} in terms of the triangle squared lengths $\sigma_t, \sigma_s$ on the boundary and $\sigma$ in the interior. Therefore
\begin{align}\label{eq:s27}
\sigma_{27} = \frac{\sigma_{12}}{\alpha_{125}^2 \alpha_{425}^2}
\end{align}
determines the light ray location in terms of $\sigma_t, \sigma_s$ and $\sigma<0$.

Similarly for a box with spacelike interior edges as in the middle of \Cref{fig:lcf_symmetry-reduced-box_light-ray},
\begin{align}\label{eq:s37}
\sigma_{37} = \sigma_{13} \alpha_{135}^2 \alpha_{435}^2
\end{align}
determines the light ray location in terms of $\sigma_t, \sigma_s$ and $\sigma>0$.

It is convenient to represent the light ray location using a dimensionless variable that grows linear with respect to length instead of squared length. We define the \textbf{dimensionless light ray location} as a function of the interior squared length $\sigma$ by
\begin{align}\label{eq:r}
r(\sigma) =
\begin{cases}
-\frac{\sqrt{\sigma_{47}}}{\sqrt{\sigma_{34}}}=\frac{\sqrt{\sigma_{37}}}{\sqrt{\sigma_{34}}}-1=\frac{\sqrt{\sigma_{13} \alpha_{135}^2 \alpha_{435}^2}}{\sqrt{\sigma_{34}}}-1, \quad &\text{spacelike interior edge } \sigma>0,
\\
0  &\text{lightlike interior edge } \sigma=0,
\\
\frac{\sqrt{\sigma_{47}}}{\sqrt{\sigma_{24}}}=1-\frac{\sqrt{\sigma_{27}}}{\sqrt{\sigma_{24}}}=1-\frac{\sqrt{\sigma_{12}\alpha_{125}^{-2} \alpha_{425}^{-2}}}{\sqrt{\sigma_{24}}}, &\text{timelike interior edge } \sigma<0.
\end{cases}
\end{align}
As the light ray location moves continuously from vertex 3 to 4 to 2, the value of $r$ grows from $-1$ to $0$ to $1$ linearly so that, for instance, $r=-0.5$ when the light ray lands right in the middle between vertices 3 and 4. 

\subsection{Light ray fluctuations}

Formula \eqref{eq:r} allows one to compute the light ray location for a given configuration. When different configurations are summed over in a path integral, there are quantum fluctuations in the light ray location.

\begin{figure}
    \centering
    \includegraphics[width=.25\textwidth]{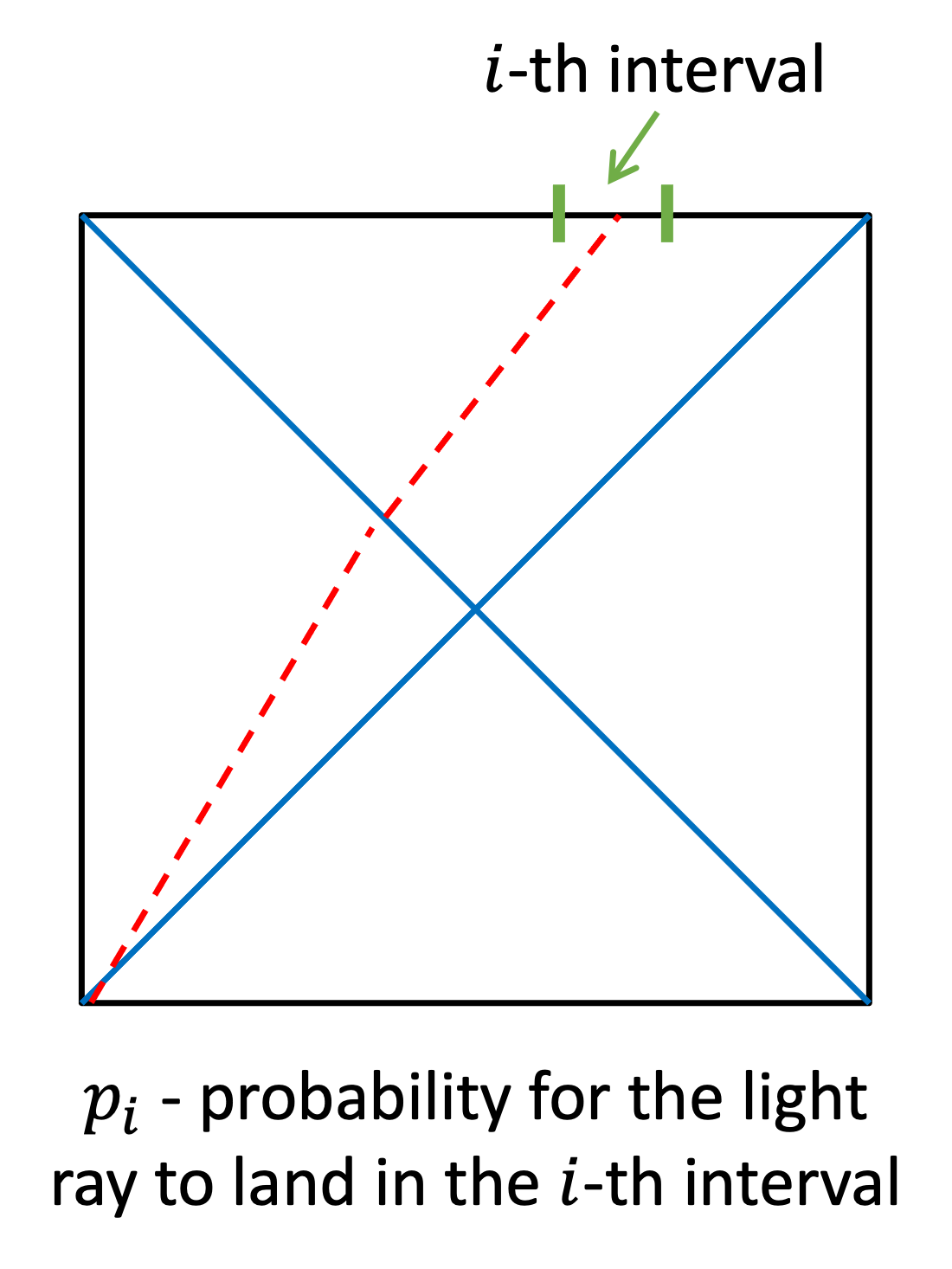}
    \caption{Partitioning the possible light ray locations into intervals and assigning relative probabilities according to the gravitational path integral.}
    \label{fig:lcf_light-ray-intervals}
\end{figure}

To characterize the quantum fluctuation of light ray location, we partition the possible values of $r$ into $N$ many equal size intervals $I_i, i = 1,2,\cdots, N$. Path integrating over the spacetime configurations compatible with the set of light ray locations $I_i$ yields the amplitude
\begin{align}\label{eq:lra}
A_i=\int_{\sigma:r(\sigma)\in I_i} d\sigma ~ L[\sigma] C[\sigma] \mu[\sigma] e^{E[\sigma]}.
\end{align}
Given $r(\sigma)$, the values of $\sigma$ corresponding to $I_i$ can be solved numerically so that the integration domain is determined. From these amplitude, one could compute the relative probability for the light ray to land in the $i$-th interval (\Cref{fig:lcf_light-ray-intervals})
\begin{align}\label{eq:pi}
p_i=\frac{\abs{A_i}^2}{\sum_i \abs{A_i}^2},
\end{align}
The probability distribution over $i$ informs us how much light ray fluctuation there is.

\section{Light ray fluctuations in 2D}\label{sec:lrf2d}

\subsection{Fixed boundary size, varying coupling constants}

In this section we fix the boundary squared lengths to $\sigma_s=1, \sigma_t=-1$ and study the light ray fluctuation for different sets of coupling constants in 2D. Varying boundary squared lengths will be considered in the next section.

\begin{figure}
    \centering
    \includegraphics[width=.35\textwidth]{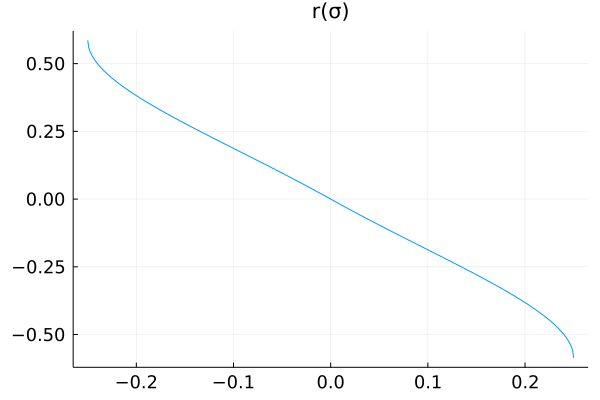}
    \caption{The light ray location $r(\sigma)$ in 2D as a function of the interior edge squared length $\sigma$ when $\sigma_s=1, \sigma_t=-1$.}
    \label{fig:lcf_rplot}
\end{figure}

The Lorentzian and lightcone constraints of \Cref{sec:llc} limit the interior edge squared length to $\sigma\in (\sigma_t/4, \sigma_s/4)= (-0.25, 0.25)$. Since the boundary values $\sigma= -0.25, 0.25$ are the branch point singularities for the path integral exponent, we take the integration domain for $\sigma$ to be $[-b,b]$ where
\begin{align}
b = 0.25-\epsilon.
\end{align}
The choice of the cutoff value $\epsilon$ does not influence the probability distributions to be computed in any significant way, as long as it is kept small. Here we take $\epsilon=10^{-8}$.

The light ray location $r(\sigma)$ as a function of the interior edge squared length is plotted in \Cref{fig:lcf_rplot} according to \eqref{eq:r}. Clearly the range of $r$ is a proper subset of $[-1,1]$, so not all locations on the boundary edges 34 and 24 are reachable by light rays emanating from vertex 1 in the current setting. We partition the \textit{reachable} light ray locations into
\begin{align}
N=16
\end{align}
equal size intervals, which are denoted $I_i$ for $i = 1,2,\cdots, 16$. 

In the following, we compute the light location probabilities $p_i$ for $i=1,2,\cdots, 16$ defined in \eqref{eq:pi} through numerical integration in the Julia programming language \cite{Bezanson2017Julia:Computing} using the QuadGK package based on the adaptive Gauss-Kronrod quadrature method.

The results are presented below in figure \Cref{fig:2D_probabilities-amplitudes_lambda} and onward. In all the figures, the coupling constants are displayed to 2 significant digits which explains the $\approx$ sign in the titles. Each bar chart shows a probability distribution of $p_i$ kept to 4 decimal places with some fixed set of parameters. The path integral amplitudes as functions of $\sigma$ are plotted below the bar charts. The measure factor $\mu[\sigma]$ is counted as a factor within the amplitude so that the integral is with respect to the plain Lebesgue measure in $\sigma$. From \Cref{fig:lcf_rplot} the light ray location is a decreasing function of $\sigma$, so the bar charts $p_i$ are shown for $i=16,\cdots,1$ from left to right in order to match the increasing values of $\sigma$ for the amplitude plots. This eases the comparison between the bar charts and the amplitude plots, and one observes that the places where the amplitude varies slowly corresponds well with the peaks of the probability distribution.



\subsubsection*{One non-vanishing parameter}

\begin{figure}
    \centering
    \includegraphics[width=1.0\textwidth]{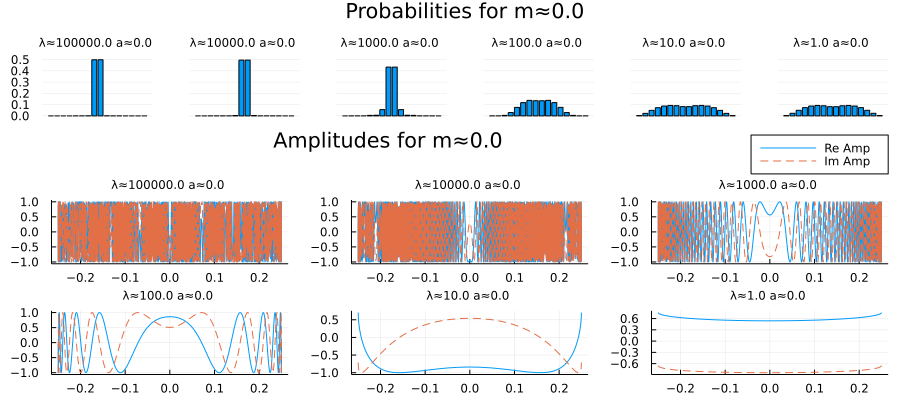}
    \caption{Probability and amplitude distributions for a family of $\lambda$ with $m=0, a=0$. By \Cref{fig:lcf_rplot} the light ray location is a decreasing function of $\sigma$. Therefore in this and the following figures of $2D$, the probabilities $p_i$ of \eqref{eq:pi} are plotted for $i=16,\cdots,1$ from left to right in order to match the increasing values of $\sigma$ for the amplitude plots.}
    \label{fig:2D_probabilities-amplitudes_lambda}
\end{figure}

\begin{figure}
    \centering
    \includegraphics[width=1.0\textwidth]{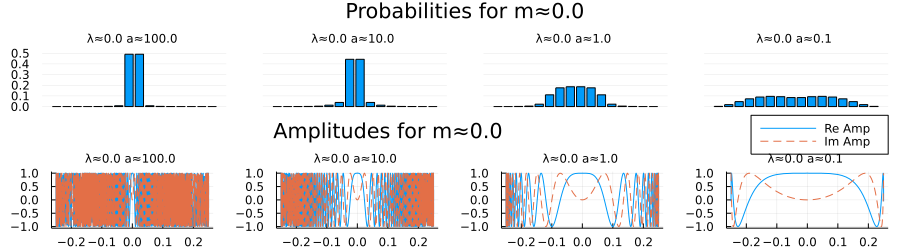}
    \caption{Probability and amplitude distributions for a family of $a$ with $m=0, \lambda=0$.}
    \label{fig:2D_probabilities-amplitudes_a}
\end{figure}

\begin{figure}
    \centering
    \includegraphics[width=1.0\textwidth]{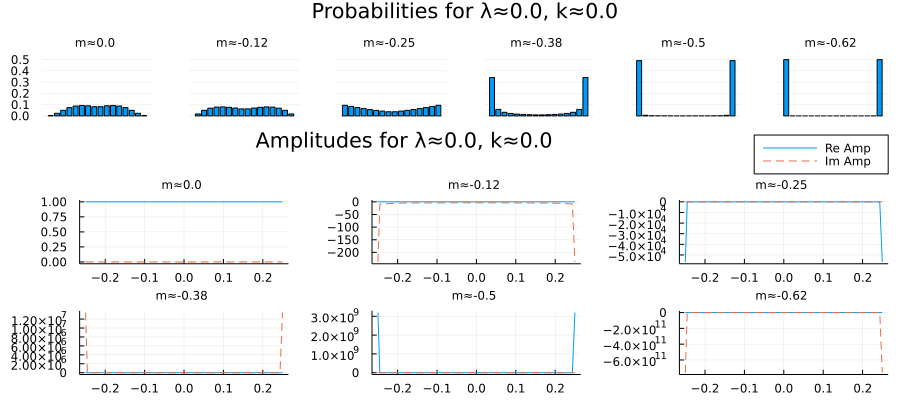}
    \caption{Probability and amplitude distributions for a family of $m$ with $\lambda=0, a=0$.}
    \label{fig:2D_probabilities-amplitudes_m}
\end{figure}

As noted in \Cref{sec:2D}, the Gauss-Bonnet theorem implies that the Einstein-Hilbert term is a constant in the present model with fixed topology. This constant term drops out as a common factor for the numerator and denominator in the definition \eqref{eq:pi} for $p_i$. Therefore there are only three non-trivial free parameters $m,\lambda,a$. We first consider the cases where only one of the three parameters is non-zero.

When a parameter $x=\lambda$ or $a$ is non-vanishing, it can be both positive and negative. However, $p_i$ for $x$ equals $p_i$ for $-x$, because for the amplitudes are complex conjugates for all configurations. For this reason we only display the data for positive $\lambda$ and $a$ in \Cref{fig:2D_probabilities-amplitudes_lambda} and \Cref{fig:2D_probabilities-amplitudes_a}.

When only $\lambda>0$ the results are shown in \Cref{fig:2D_probabilities-amplitudes_lambda}. For large $\lambda\gtrsim 1000$ light ray fluctuation is small as $p_i$ is sharply peaked. As $\lambda$ gets smaller the amount of fluctuation gets larger. A similar conclusion holds when only $a> 0$ (\Cref{fig:2D_probabilities-amplitudes_a}), namely light ray fluctuation is small and large respectively when $a$ is large and small. The reason is clear from the amplitude plots below the probabilities. A large parameter makes the phase of the integral amplitude change faster so that probabilities are suppressed except for special regions where the phase is nearly stationary.

These fit the common intuition that as $\hbar$ gets smaller quantum fluctuations become smaller. The path integral exponent $E$ of \eqref{eq:E2D} is related to the action $S$ by $E=\frac{i}{\hbar}S$, so the coupling constants $\lambda$ and $a$ scale inverse-proportionally with $\hbar$. Therefore a smaller $\hbar$ means a large absolute value for $\lambda$ and $a$, and we saw that these yield smaller light ray fluctuations.

Next we consider $m<0$ with the results shown in \Cref{fig:2D_probabilities-amplitudes_m}. The $p_i$ values start distributed evenly for $m=0$, and gets pushed towards the two sides as $m$ decreases. This trend is easy understand. In the definition \eqref{eq:mf} $\sV_s$ is always negative so for the probabilities only $\abs{\sV_s}$ matters. The smaller the negative exponent $m$ is, the more it favours configurations with small $\abs{\sV_s}$, which in the current model means larger values of $\abs{\sigma}$ according to \eqref{eq:2dsvol}. Therefore for a very small negative $m$ the light ray location is concentrated around the two sides of the plots.

In summary if $x$ is the only non-vanishing parameter among the three parameters, large $\abs{x}$ suppresses light ray fluctuations while small $\abs{x}$ enhances light ray fluctuations.


\subsubsection*{Multiple non-vanishing parameters}

When multiple parameters are non-vanishing, one could expect that the qualitative features of the probability distribution follows that of the dominating parameter. This indeed holds in the cases studied below.

\begin{figure}
    \centering
    \includegraphics[width=1.0\textwidth]{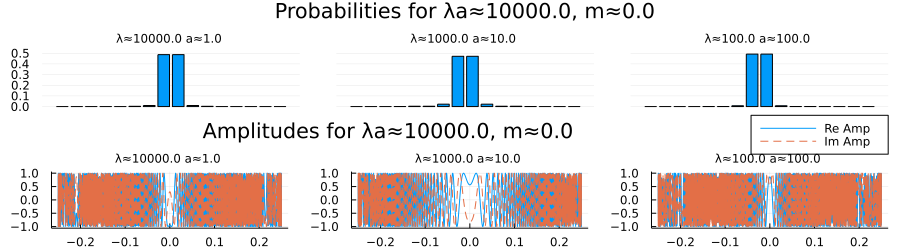}
    \caption{Probability and amplitude distributions for a family of $(\lambda,a)$ with $\lambda a=10^4, m=0$.}
    \label{fig:2D_probabilities-amplitudes_lambdaa1}
\end{figure}

\begin{figure}
    \centering
    \includegraphics[width=1.0\textwidth]{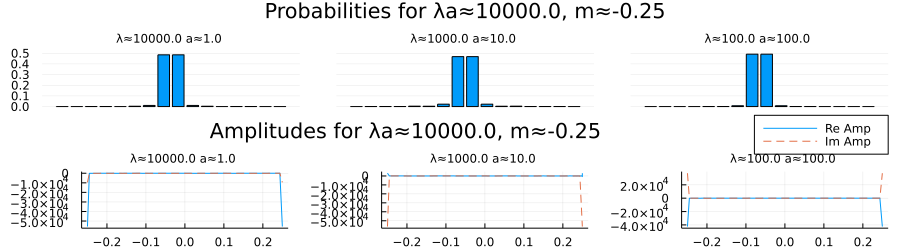}
    \caption{Probability and amplitude distributions for a family of $(\lambda,a)$ with $\lambda a=10^4, m=-0.25$.}
    \label{fig:2D_probabilities-amplitudes_lambdaa2}
\end{figure}

For reasons that will become clear in \Cref{sec:fccvbs}, we organize the set of parameters according to the product $\lambda a$. We consider $m=0$ and $m=-0.25$ (the value for the Dewitt measure \cite{Hamber2009QuantumApproach}) but not smaller values as they only serve to push the probability distribution towards the two sides like in \Cref{fig:2D_probabilities-amplitudes_m}.

The results for $\lambda a=10^4$ are shown in \Cref{fig:2D_probabilities-amplitudes_lambdaa1} and \Cref{fig:2D_probabilities-amplitudes_lambdaa2}. Light ray fluctuation is small for all the 6 families of parameters shown. Here with $\lambda a=10^4$, there is always at least one of $\lambda$ and $a$ that is large to yield fast oscillation of the phase and suppress the probabilities away from the center region.

\begin{figure}
    \centering
    \includegraphics[width=1.0\textwidth]{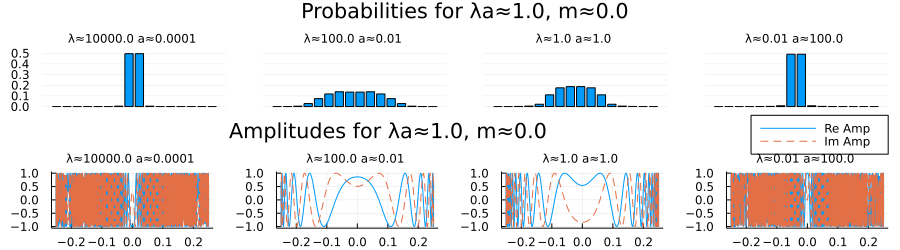}
    \caption{Probability and amplitude distributions for a family of $(\lambda,a)$ with $\lambda a=1, m=0$.}
    \label{fig:2D_probabilities-amplitudes_lambdaa=1}
\end{figure}

\begin{figure}
    \centering
    \includegraphics[width=1.0\textwidth]{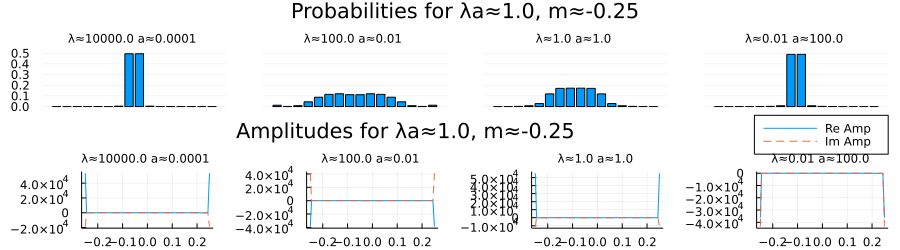}
    \caption{Probability and amplitude distributions for a family of $(\lambda,a)$ with $\lambda a=1, m=-0.25$.}
    \label{fig:2D_probabilities-amplitudes_lambdaa=1_}
\end{figure}

The results for $\lambda a=1$ is shown in \Cref{fig:2D_probabilities-amplitudes_lambdaa=1} and \Cref{fig:2D_probabilities-amplitudes_lambdaa=1_}. For this smaller value of $\lambda a$ as well, the probabilities are concentrated around the center when $\lambda\gtrsim 1000$ or $a\gtrsim 100$. Different from the case of $\lambda a=10^4$, there are now families of parameters in which both $\lambda$ and $a$ are small (e.g., $\lambda=1, a=1$) so that light ray fluctuation is larger.

\begin{figure}
    \centering
    \includegraphics[width=1.0\textwidth]{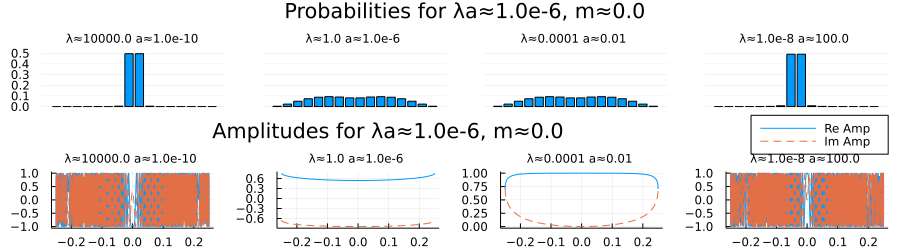}
    \caption{Probability and amplitude distributions for a family of $(\lambda,a)$ with $\lambda a=10^{-6}, m=0$.}
    \label{fig:2D_probabilities-amplitudes_lambdaa=1e-6}
\end{figure}

\begin{figure}
    \centering
    \includegraphics[width=1.0\textwidth]{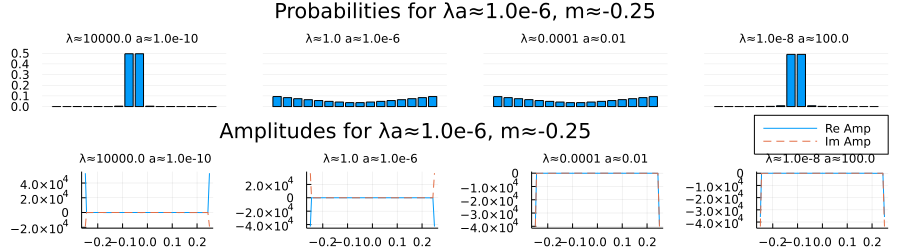}
    \caption{Probability and amplitude distributions for a family of $(\lambda,a)$ with $\lambda a=10^{-6}, m=-0.25$.}
    \label{fig:2D_probabilities-amplitudes_lambdaa=1e-6_}
\end{figure}

As shown in \Cref{fig:2D_probabilities-amplitudes_lambdaa=1e-6} and \Cref{fig:2D_probabilities-amplitudes_lambdaa=1e-6_}, the results for a yet smaller positive value $\lambda a=10^{-6}$ exhibits the same qualitative features, which we expect to hold generically for small positive values of $\lambda a$.

\begin{figure}
    \centering
    \includegraphics[width=1.0\textwidth]{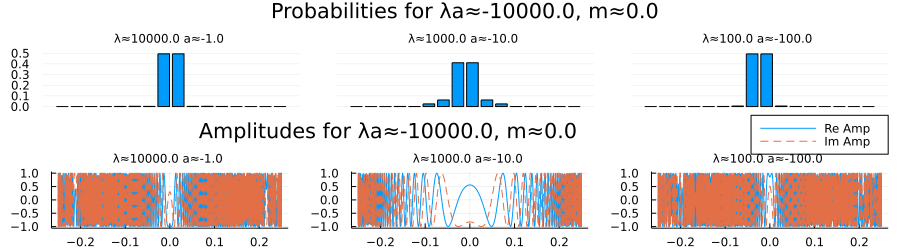}
    \caption{Probability and amplitude distributions for a family of $(\lambda,a)$ with $\lambda a=-10^4, m=0$.}
    \label{fig:2D_probabilities-amplitudes_lambdaa1-}
\end{figure}

\begin{figure}
    \centering
    \includegraphics[width=1.0\textwidth]{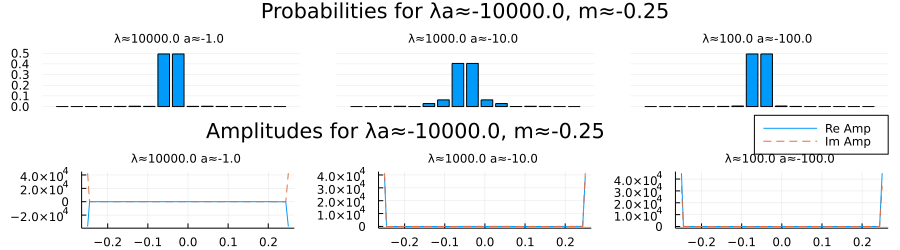}
    \caption{Probability and amplitude distributions for a family of $(\lambda,a)$ with $\lambda a=-10^4, m=-0.25$.}
    \label{fig:2D_probabilities-amplitudes_lambdaa2-}
\end{figure}

With two or more non-vanishing parameters it makes a difference to allow for negative values of $\lambda a$. The results for $\lambda a=-10^{4}$ is shown in \Cref{fig:2D_probabilities-amplitudes_lambdaa1-} and \Cref{fig:2D_probabilities-amplitudes_lambdaa2-}. In comparison with $\lambda a=10^{4}$, the only major difference is that for $\lambda=1000$, light ray fluctuation is less suppressed. A comparison of the amplitude plots show that the negative $a=-10$ induces a wider region of slowly changing phases, which explains the difference in probabilities.

We also studied other negative values of $\lambda a$ opposite to the positive ones studied above. The plots turned out to be similar to the corresponding positive values so are not shown.

In summary, when $\lambda,a\ne 0$, the presence of a large $\abs{\lambda}$ or $\abs{a}$ suppresses light ray fluctuations while their absence enhances light ray fluctuations. In special cases, changing a parameter with smaller magnitude (e.g., $\lambda=1000,a=10$ to $\lambda=1000,a=-10$) can mildly affect the amount of light ray fluctuation due to cancellations between terms in the exponent.

\subsection{Fixed coupling constants, varying boundary sizes}\label{sec:fccvbs}

To study how light ray fluctuations are affected by the size of the region specified by the boundary edge lengths, we can exploit the scaling identity of \Cref{sec:si}. In 2D \eqref{eq:si2} implies that
\begin{align}
Z[l^2\sigma_B,\lambda,k,a,m] =&l^{2N_e+4mN_t} ~ Z[\sigma_B,l^2\lambda,k,l^{-2}a,m].
\end{align}
Since the light ray location $r$ is a dimensionless quantity, one could easily check that the equation holds when $Z$ is replaced by $A_i$ of \eqref{eq:lra}. Consequently
\begin{align}\label{eq:spi}
p_i[l^2\sigma_B,\lambda,k,a,m] = p_i[\sigma_B,l^2\lambda,k,l^{-2}a,m].
\end{align}

This says that rescaling the boundary condition (LHS) is equiavlent to rescaling the parameters $\lambda$ and $a$ while keeping $\lambda a$ fixed. Therefore the previous results for fixed $\lambda a$ also inform us about how light ray fluctuations depend on the size of the region.
For instance, suppose we fix the parameters to be $\lambda=10^4, a=1, m=0$. Then for $\sigma_s=-\sigma_t=1$, $p_i$ are as in the first plot in \Cref{fig:2D_probabilities-amplitudes_lambdaa1}. The next plots are for the boundary squared lengths shrinked by factors of $l^2=0.1$.

All the previous figures for fixed $\lambda a$ can then be read from left to right as a progressive shrinking of the size of the boundary for fixed parameters. We see that light ray fluctuation is suppressed for both large and small sized boundaries.

\subsection{Time-space duality}

A notable feature of the data for $p_i$ is that the values of $p_i$ are completely symmetrical in the sense that $p_i=p_{N-i}$ for $i=1,2,\cdots, N/2$. This is not a coincidence but follows from some symmetry considerations.

Given a path integral configuration $\sigma$, consider the map $\sigma\mapsto -\sigma$ that negates all the squared lengths including the boundary ones. Physically, this map exchanges spacelike intervals and timelike intervals. It was noted that this map is a symmetry for 2D Lorentzian quantum gravity with the cosmological constant term and the Einstein-Hilbert term \cite{Jia2022Time-spaceGravity}.

Here we note that the symmetry holds more generally for any theory with additional even powers of the Ricci scalar $R$ in the action. According to \Cref{sec:2D}, the simplicial analogue of $R$ is $-\delta_v/A_v$. From the definitions of $\delta_v$ and $A_v$, $\sigma\mapsto -\sigma$ maps $\delta_v$ to $-\delta_v$ and $A_v$ to $A_v$. Hence $-\delta_v/A_v$ is mapped to $\delta_v/A_v$, and any even power of $-\delta_v/A_v$ is left invariant. Since the map $g\mapsto -g$ on the metric field $g$ also takes $R$ to $-R$, the symmetry is also expected to hold for additional theories of 2D quantum gravity beyond simplicial quantum gravity.

\begin{figure}
    \centering
    \includegraphics[width=.6\textwidth]{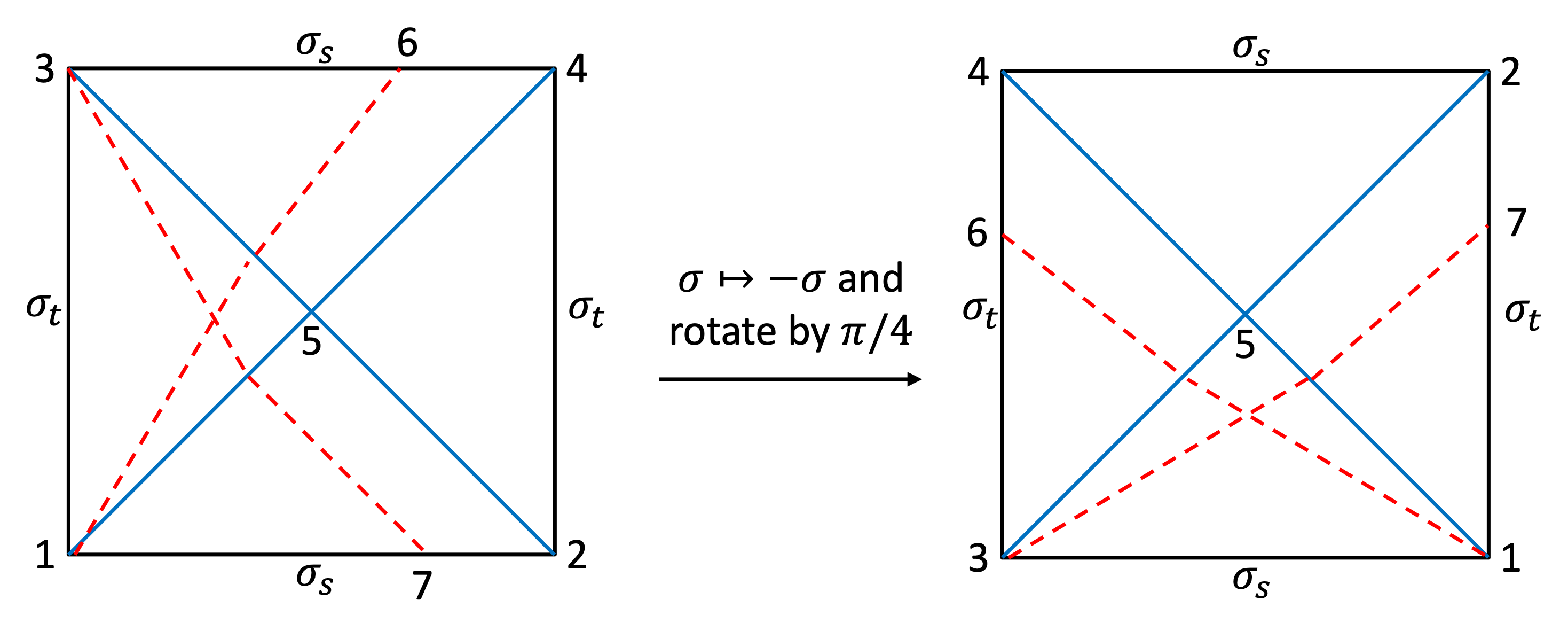}
    \caption{Applying $\sigma\mapsto -\sigma$ followed by a $\pi/4$ rotation to a box configuration obeying $\sigma_t=-\sigma_s$ yields a box configuration obeying the same boundary condition. The light ray location $r$ at $6$ is changed to $7$, which by the symmetry of the model is located at $-r$.}
    \label{fig:lcf_tsd}
\end{figure}

While the map $\sigma\mapsto -\sigma$ exchanges timelike and spacelike intervals, it preserves lightlike intervals. Applied to a box model configuration it turns the timelike boundaries to spacelike ones and \textit{vice versa}, while preserving all light ray paths. For the models studied the boundary condition obeys $\sigma_t=-\sigma_s$, so a rotation of the box by $\pi/4$ in either direction yields back a configuration obeying the same boundary condition. On the other hand the light ray location is changed from $r$ to $-r$ \Cref{fig:lcf_tsd}. Since the map $\sigma\mapsto -\sigma$ preserves path integral amplitudes, we see that each box configuration with light ray location $r$ corresponds to another configuration of the same amplitude with light ray location $-r$. In the 2D symmetry-reduced model $r\in I_i$ if and only if $-r\in I_{N-i}$, so $A_i=A_{N-i}$ and $p_i=p_{N-i}$.

\section{Light ray fluctuations in 3D}\label{sec:lrf3d}

\subsection{Fixed boundary size, varying coupling constants}

\begin{figure}
    \centering
    \includegraphics[width=.35\textwidth]{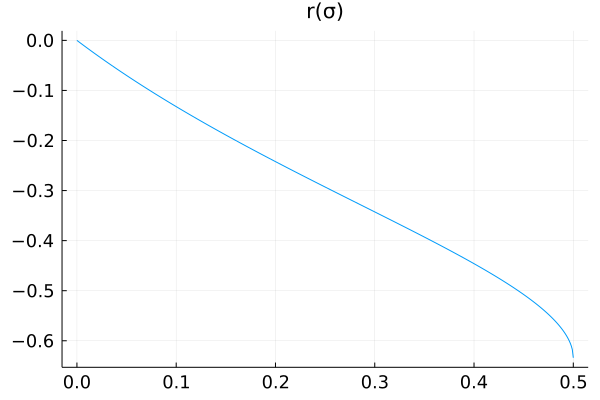}
    \caption{The light ray location $r(\sigma)$ in 3D as a function of the interior edge squared length $\sigma$ when $\sigma_s=1, \sigma_t=-1$.}
    \label{fig:lcf_rplot3D}
\end{figure}

The study in 3D is parallel to the case of 2D. As illustrated in \Cref{fig:lcf_symmetry-reduced}, the light ray travels in the diagonal plane. Again we fix the boundary squared lengths to $\sigma_s=1, \sigma_t=-1$. The Lorentzian and lightcone constraints limit the interior edge squared length to $\sigma\in ((\sigma_s+\sigma_t)/4, \sigma_s/2)=(0.0,0.5)$. In particular, four light rays emanate from the interior vertex in the diagonal plane if and only if two lightcones are attached to the interior vertex in the 3D box, so this condition is enough to ensure the lightcone constraint is obeyed. Like in 2D, the integration domain is cutoff with $\epsilon=10^{-8}$ at
\begin{align}
(\epsilon,0.5-\epsilon)
\end{align}
to avoid branch point singularities in the domain. The choice of the cutoff value $\epsilon$ does not influence the probability distributions to be computed in any significant way.

In computing the probabilities through numerical integration we need to impose an additional cutoff around $\sigma=\sigma_s/4$. At this point some angles evaluated according to \eqref{eq:theta} diverge, but the results should be finite since the divergences of different angles cancel in path integral exponent. However to avoid numerical inaccuracies due to the difference of large numbers, a cutoff is imposed. Where $\abs{\sigma-\sigma_s/4}<\epsilon'=10^{-7}$ in the integration domain, the integrand is evaluated at $\sigma=\sigma_s/4\pm \epsilon'$ for $\sigma>,<\sigma_s/4$. This value of $\epsilon'$ is picked to be small without leading to numerical infinities.

The light ray location $r(\sigma)$ as a function of the interior edge squared length is plotted in \Cref{fig:lcf_rplot3D} according to \eqref{eq:r} when $\sigma_{12}=\sigma_{34}=2\sigma_s=2$ as illustrated in \Cref{fig:lcf_symmetry-reduced}. The reachable light ray locations are again partitioned into
\begin{align}
N=16
\end{align}
equal size intervals, which are denoted $I_i$ for $i = 1,2,\cdots, 16$.

In 3D the parameters are $\lambda,k,m$ for the cosmological constant term, the Einstein-Hilbert term, and the measure term. In figures such as \Cref{fig:3D_probabilities-amplitudes_lambda} the probability distributions of $p_i$ to 4 decimal places are plotted in the bar charts for $i=16,\cdots,1$ from left to right to match the increasing values of $\sigma$ for the amplitude plots. The range of probability plots is limited to $[0.0,0.5]$ to make the low probability bars more visible.

\subsubsection*{One non-vanishing parameter}

\begin{figure}
    \centering
    \includegraphics[width=1.0\textwidth]{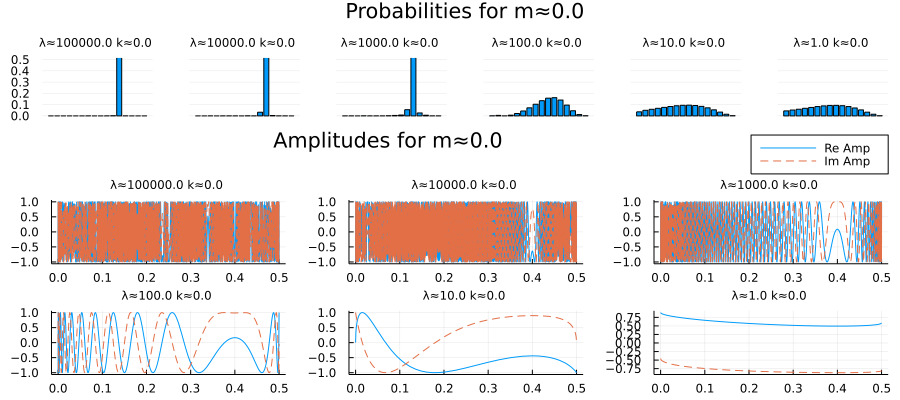}
    \caption{Probability and amplitude distributions for a family of $\lambda$ with $m=0, k=0$. By \Cref{fig:lcf_rplot3D} the light ray location is a decreasing function of $\sigma$. Therefore in this and the following figures of $3D$, the probabilities $p_i$ of \eqref{eq:pi} are plotted for $i=16,\cdots,1$ from left to right in order to match the increasing values of $\sigma$ for the amplitude plots.}
    \label{fig:3D_probabilities-amplitudes_lambda}
\end{figure}

\begin{figure}
    \centering
    \includegraphics[width=1.0\textwidth]{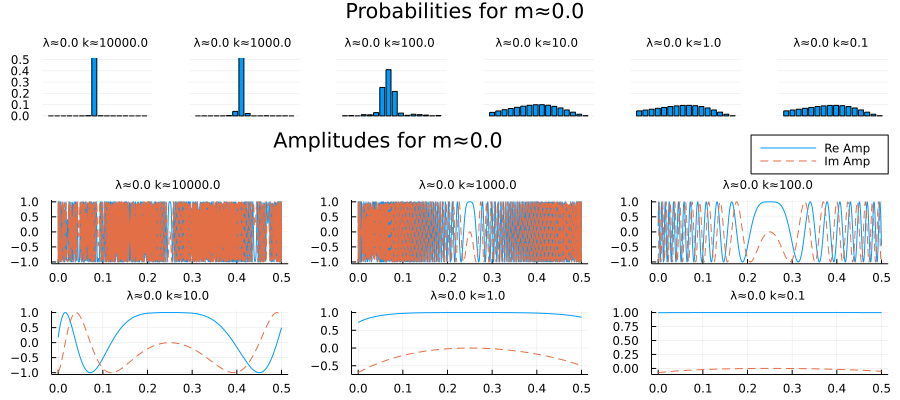}
    \caption{Probability and amplitude distributions for a family of $a$ with $m=0, \lambda=0$.}
    \label{fig:3D_probabilities-amplitudes_a}
\end{figure}

\begin{figure}
    \centering
    \includegraphics[width=1.0\textwidth]{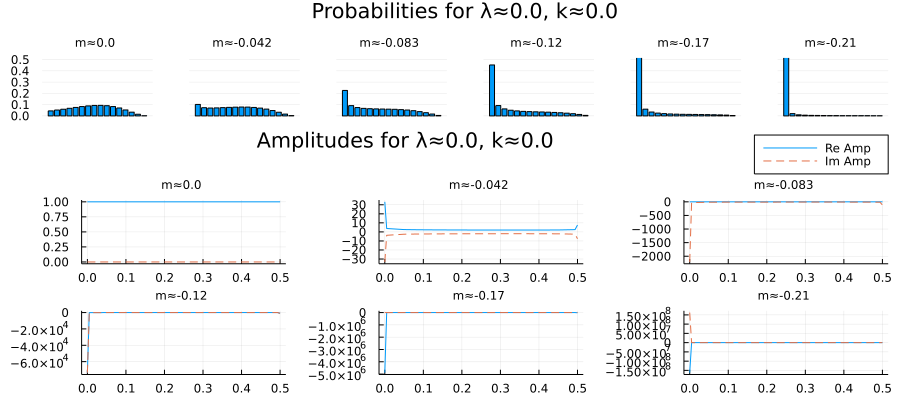}
    \caption{Probability and amplitude distributions for a family of $m$ with $\lambda=0, k=0$.}
    \label{fig:3D_probabilities-amplitudes_m}
\end{figure}

We first consider the cases where only one of the three parameters is non-zero. We only display the results for positive $\lambda$ and $k$ in \Cref{fig:3D_probabilities-amplitudes_lambda} and \Cref{fig:3D_probabilities-amplitudes_a}, because as in the 2D case $p_i$ for the positive and negative parameters are equal.

For large $\lambda\gtrsim 1000$ or $k\gtrsim 1000$, $p_i$ is sharply peaked indicating small light ray fluctuations. As $\lambda$ or $k$ gets smaller the fluctuation gets larger.
The reason is clear from the amplitude plots below the probabilities. As in the 2D case, the large parameter makes the phase of the integral amplitude change faster so that probabilities are suppressed except for special regions where the phase is nearly stationary.
As in the 2D case, these fit the common intuition that as $\hbar$ gets smaller quantum fluctuations become smaller. The path integral exponent $E$ of \eqref{eq:EgD} is related to the action $S$ by $E=\frac{i}{\hbar}S$, so the coupling constants $\lambda$ and $k$ scale inverse-proportionally with $\hbar$. Therefore a smaller $\hbar$ means a larger absolute value for $\lambda$ and $a$, and we see that these yield smaller light ray fluctuations.

The results for $m< 0$ are shown in \Cref{fig:3D_probabilities-amplitudes_m}. The $p_i$ values start distributed evenly for $m=0$, and gets pushed towards the left side as $m$ decreases. Like in 2D, the smaller the negative parameter $m$ is, the more it favours configurations with small $\abs{\sV_s}$. In 3D this favours smaller values of $\sigma$ according to \eqref{eq:svol} and explains why the light ray location is concentrated around the left side of the plots.

In summary if $x$ is the only non-vanishing parameter among the three parameters, large $\abs{x}$ suppresses light ray fluctuations while small $\abs{x}$ enhances light ray fluctuations.

\subsubsection*{Multiple non-vanishing parameters}

\begin{figure}
    \centering
    \includegraphics[width=1.0\textwidth]{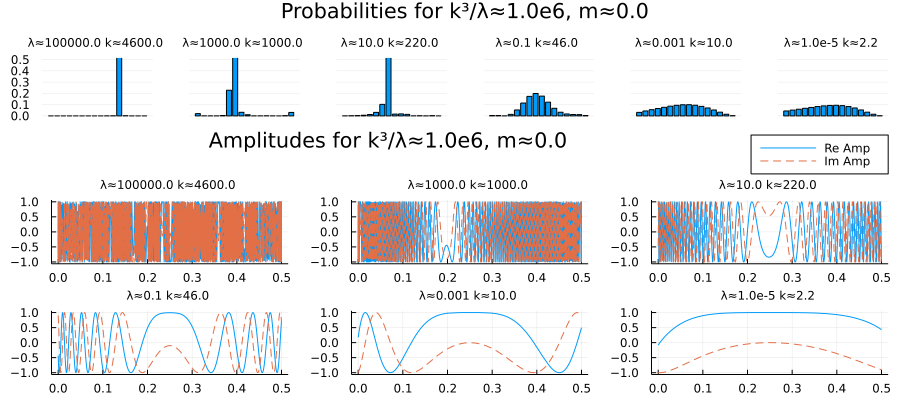}
    \caption{Probability and amplitude distributions for a family of $(\lambda,k)$ with $k^3/\lambda =10^6, m=0$.}
    \label{fig:3D_probabilities-amplitudes_lambdaa1}
\end{figure}

\begin{figure}
    \centering
    \includegraphics[width=1.0\textwidth]{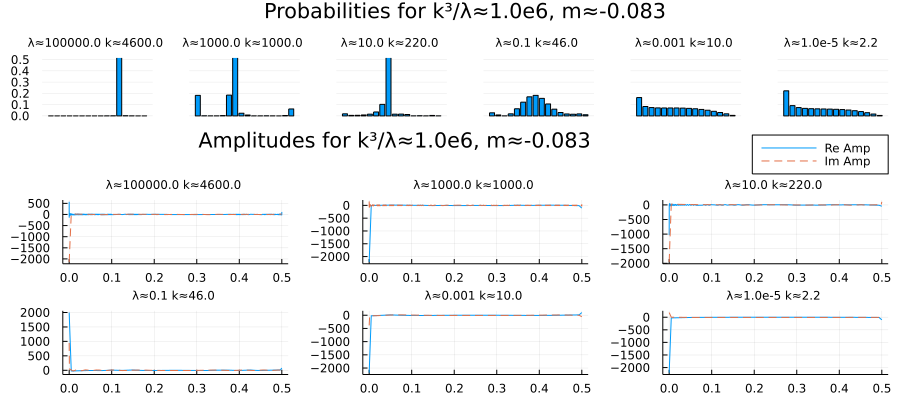}
    \caption{Probability and amplitude distributions for a family of $(\lambda,k)$ with $k^3/\lambda =10^6, m\approx-1/12$.}
    \label{fig:3D_probabilities-amplitudes_lambdaa2}
\end{figure}


In 3D the scaling analysis requires fixing $k^3/\lambda$ (\Cref{eq:3dvbs}), so the parameters are organized according to the value of $k^3/\lambda$. We consider $m=0$ and $m=-1/12$ (the value for the Dewitt measure \cite{Hamber2009QuantumApproach}) but not smaller values as they only serve to push the probability distribution towards the side like in \Cref{fig:3D_probabilities-amplitudes_m}.

\begin{figure}
    \centering
    \includegraphics[width=1.0\textwidth]{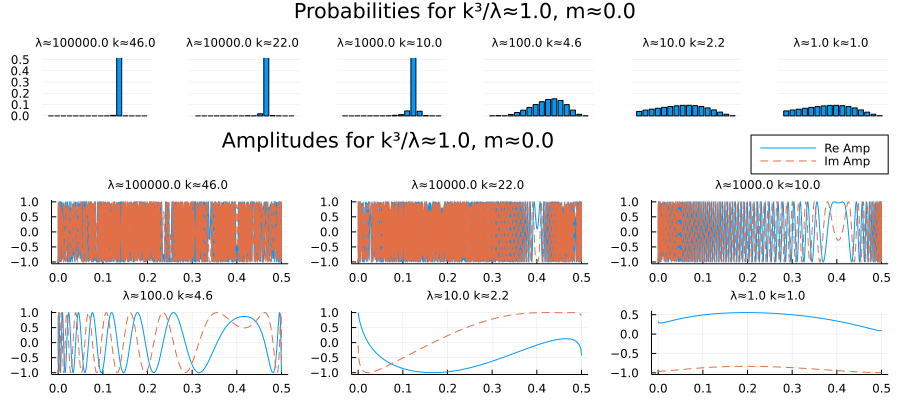}
    \caption{Probability and amplitude distributions for a family of $(\lambda,k)$ with $k^3/\lambda =1, m=0$.}
    \label{fig:3D_probabilities-amplitudes_lambdaa=1}
\end{figure}

\begin{figure}
    \centering
    \includegraphics[width=1.0\textwidth]{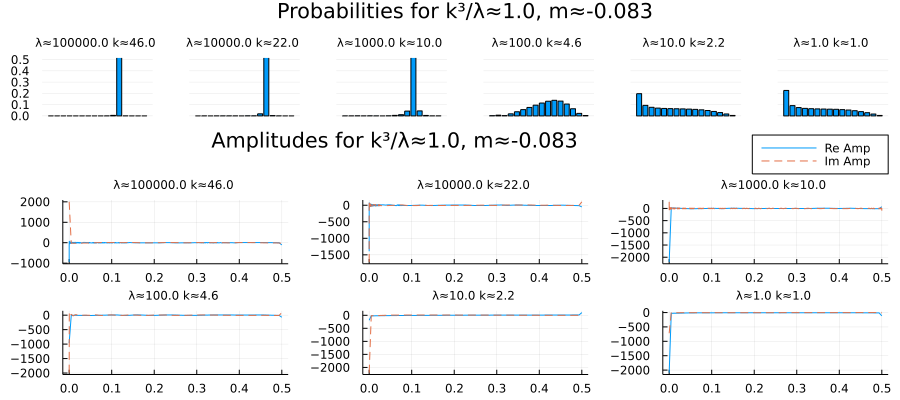}
    \caption{Probability and amplitude distributions for a family of $(\lambda,k)$ with $k^3/\lambda =1, m\approx-1/12$.}
    \label{fig:3D_probabilities-amplitudes_lambdaa=1_}
\end{figure}

\begin{figure}
    \centering
    \includegraphics[width=1.0\textwidth]{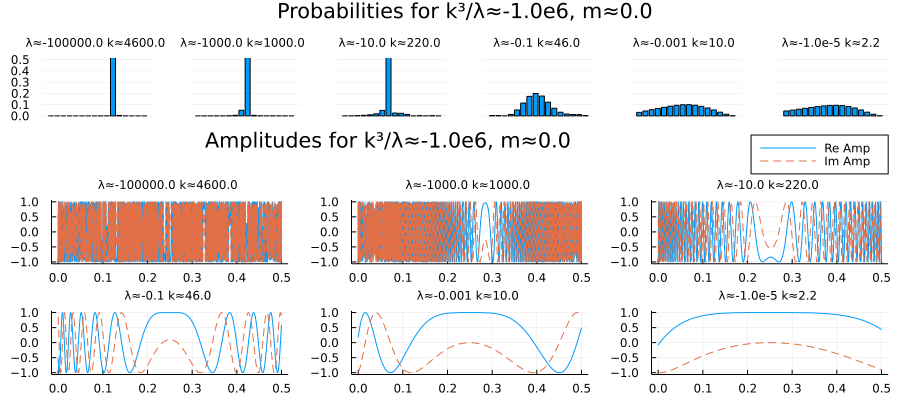}
    \caption{Probability and amplitude distributions for a family of $(\lambda,k)$ with $k^3/\lambda =-10^6, m=0$.}
    \label{fig:3D_probabilities-amplitudes_lambdaa=-1e-6}
\end{figure}

\begin{figure}
    \centering
    \includegraphics[width=1.0\textwidth]{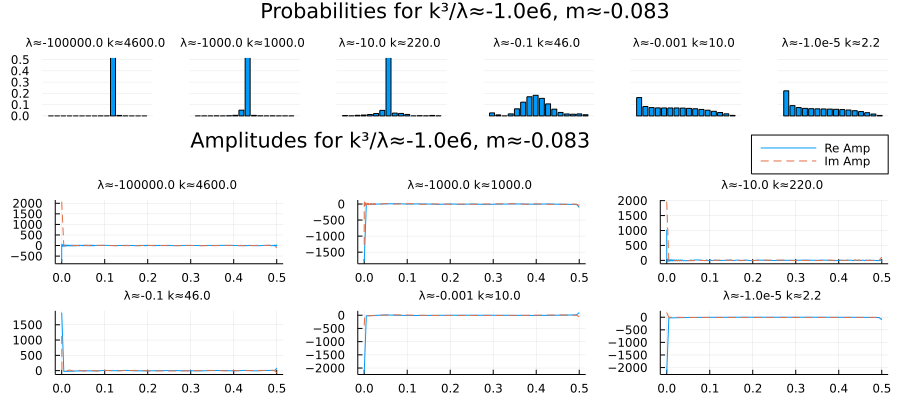}
    \caption{Probability and amplitude distributions for a family of $(\lambda,k)$ with $k^3/\lambda =-10^6, m\approx-1/12$.}
    \label{fig:3D_probabilities-amplitudes_lambdaa=-1e-6_}
\end{figure}

\begin{figure}
    \centering
    \includegraphics[width=1.0\textwidth]{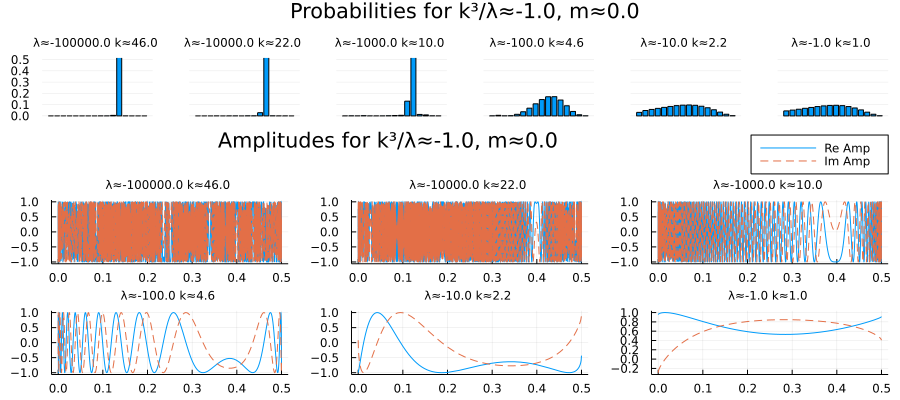}
    \caption{Probability and amplitude distributions for a family of $(\lambda,k)$ with $k^3/\lambda =-1, m=0$.}
    \label{fig:3D_probabilities-amplitudes_lambdaa1-}
\end{figure}

\begin{figure}
    \centering
    \includegraphics[width=1.0\textwidth]{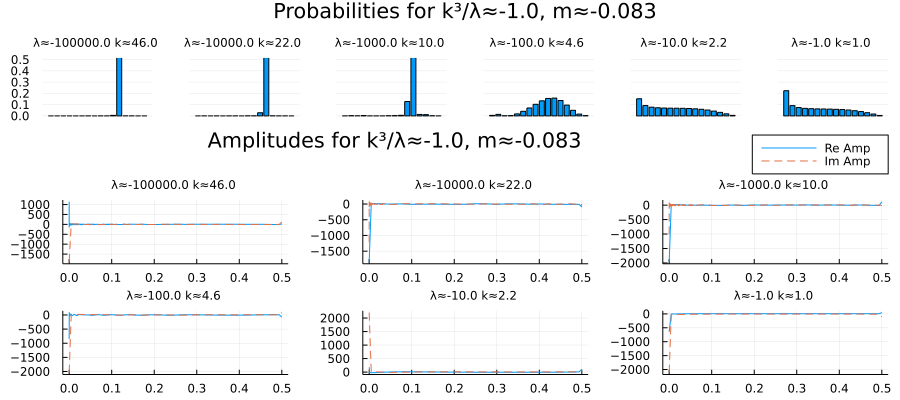}
    \caption{Probability and amplitude distributions for a family of $(\lambda,k)$ with $k^3/\lambda =-1, m\approx-1/12$.}
    \label{fig:3D_probabilities-amplitudes_lambdaa2-}
\end{figure}

The results for $k^3/\lambda=10^6$  are shown in \Cref{fig:3D_probabilities-amplitudes_lambdaa1} and \Cref{fig:3D_probabilities-amplitudes_lambdaa2}. For fixed $k^3/\lambda>0$, $\lambda$ is large when $k$ is large. The plots show that light ray fluctuation is large for large values of $\lambda$ and $k$, and small for small values of $\lambda$ and $k$. This is due to larger values of the parameters inducing fast oscillations for the phase away from the nearly stationary phase region. The results for other values of $k^3/\lambda$ including negative ones are shown in \Cref{fig:3D_probabilities-amplitudes_lambdaa=1} to \Cref{fig:3D_probabilities-amplitudes_lambdaa2-} and exhibit no qualitative difference.

The situation is contrasted with 2D where as $\lambda$ or $a$ is decreased the light ray fluctuation starts large and ends large. Clearly this is because in 2D $\lambda a$ is dimensionless and fixed so $\lambda$ and $a$ are inversely proportional, whereas in 3D $k^3/\lambda>0$ is dimensionless and fixed so $\lambda$ and $k^3$ are proportional.

\subsection{Fixed coupling constants, varying boundary sizes}\label{eq:3dvbs}

In 3D the scaling identity \eqref{eq:si2} implies that
\begin{align}
Z[l^2\sigma_B,\lambda,k,m] =&l^{2N_e+4mN_t} ~ Z[\sigma_B,l^3\lambda,lk,m],\\
p_i[l^2\sigma_B,\lambda,k,a,m] =& p_i[\sigma_B,l^3\lambda,l k,m].
\end{align}
Therefore all the previous figures for fixed $k^3/\lambda$ can then be read from left to right as a progressive shrinking of the size of the boundary for fixed parameters. We see that light ray fluctuation increases as the boundary shrinks.

\section{Light ray fluctuations in 4D}\label{sec:lrf4d}

\subsection{Fixed boundary size, varying coupling constants}

\begin{figure}
    \centering
    \includegraphics[width=.35\textwidth]{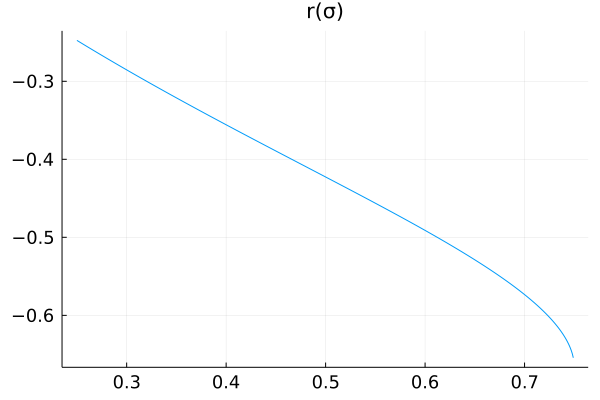}
    \caption{The light ray location $r(\sigma)$ in 4D as a function of the interior edge squared length $\sigma$ when $\sigma_s=1, \sigma_t=-1$.}
    \label{fig:lcf_rplot4D}
\end{figure}

The setting of 4D is similar to the previous ones. Again we fix the boundary squared lengths to $\sigma_s=1, \sigma_t=-1$ and the light ray travels in the diagonal plane as illustrated in \Cref{fig:lcf_symmetry-reduced}. The Lorentzian and lightcone constraints limit the interior edge squared length to $\sigma\in (\sigma_s/2+\sigma_t/4, 3\sigma_s/4=(0.25,0.75)$, and the integration domain is cutoff with $\epsilon=10^{-8}$ at
\begin{align}
(0.25+\epsilon,0.75-\epsilon).
\end{align}
In numerical integration, an additional cutoff is imposed to evaluate the integrand at $\sigma=\sigma_s/2\pm \epsilon'$ for $\sigma>,<\sigma_s/2$ when $\abs{\sigma-\sigma_s/4}<\epsilon'=10^{-7}$.
The light ray location $r(\sigma)$ as a function of the interior edge squared length is plotted in \Cref{fig:lcf_rplot4D} according to \eqref{eq:r} when $\sigma_{12}=\sigma_{34}=3\sigma_s=3$. The reachable light ray locations are again partitioned into $N=16$ equal size intervals, which are denoted $I_i$ for $i = 1,2,\cdots, 16$.
The parameters are $\lambda,k,m$ for the cosmological constant term, the Einstein-Hilbert term, and the measure term.

The probabilities and amplitudes are plotted in figures \Cref{fig:4D_probabilities-amplitudes_lambda} to \Cref{fig:4D_probabilities-amplitudes_lambdaa-1e6}. Again, the probabilities $p_i$ are kept to 4 decimal places, plotted in bars with the sequence $i=16,\cdots,1$ to match the increasing values of $\sigma$ for the amplitude plots, and the plot range is limited to $[0.0,0.5]$ to make the low probability bars more visible. The results are very similar to 3D, so we just briefly summarize them.

\subsubsection*{One non-vanishing parameter}

\begin{figure}
    \centering
    \includegraphics[width=1.0\textwidth]{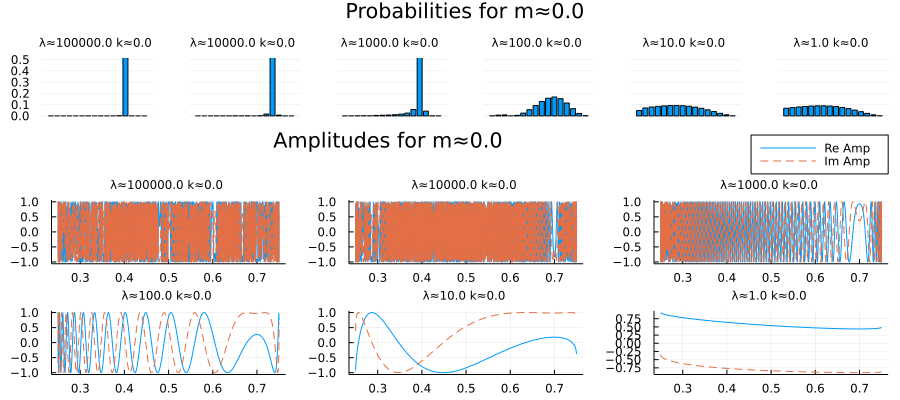}
    \caption{Probability and amplitude distributions for a family of $\lambda$ with $m=0, a=0$.  By \Cref{fig:lcf_rplot4D} the light ray location is a decreasing function of $\sigma$. Therefore in this and the following figures of $4D$, the probabilities $p_i$ of \eqref{eq:pi} are plotted for $i=16,\cdots,1$ from left to right in order to match the increasing values of $\sigma$ for the amplitude plots.}
    \label{fig:4D_probabilities-amplitudes_lambda}
\end{figure}

\begin{figure}
    \centering
    \includegraphics[width=1.0\textwidth]{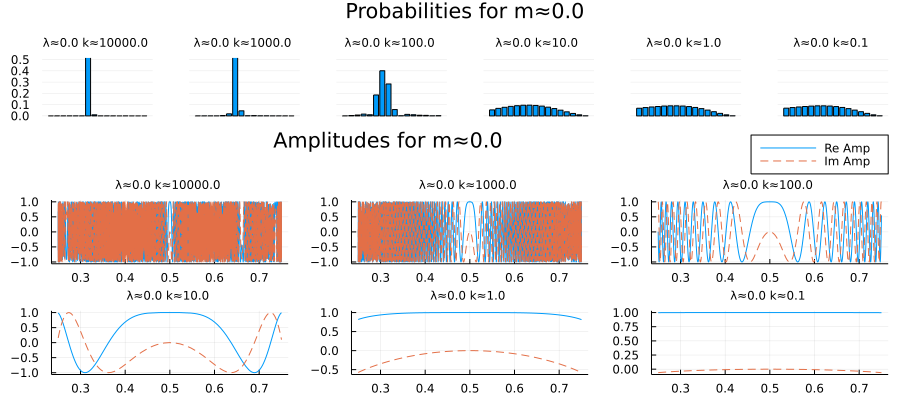}
    \caption{Probability and amplitude distributions for a family of $a$ with $m=0, \lambda=0$.}
    \label{fig:4D_probabilities-amplitudes_a}
\end{figure}

\begin{figure}
    \centering
    \includegraphics[width=1.0\textwidth]{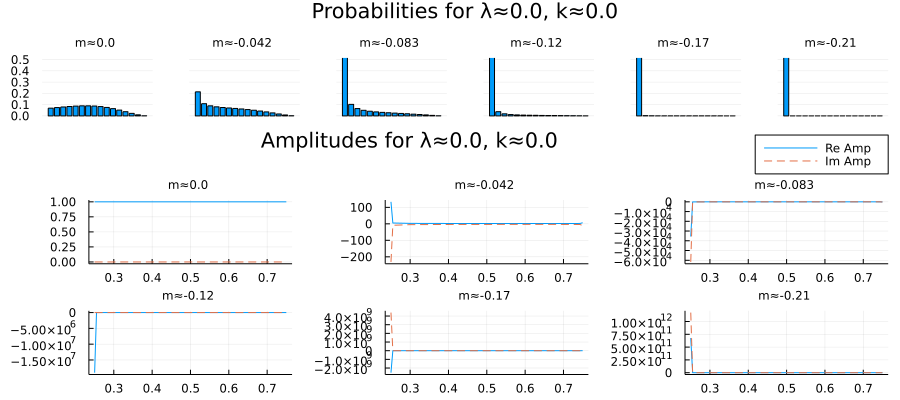}
    \caption{Probability and amplitude distributions for a family of $m$ with $\lambda=0, a=0$.}
    \label{fig:4D_probabilities-amplitudes_m}
\end{figure}

The results when only one parameter $x$ of the three parameters is non-zero are plotted in \Cref{fig:4D_probabilities-amplitudes_lambda} to \Cref{fig:4D_probabilities-amplitudes_m}. Similar to the 3D cases, large $\abs{x}$ suppresses light ray fluctuations while small $\abs{x}$ enhances light ray fluctuations. These fit the common intuition that as $\hbar$ gets smaller quantum fluctuations become smaller, since according to \eqref{eq:EgD} and $E=\frac{i}{\hbar}S$, $\abs{x}$ scale inverse-proportionally with $\hbar$, and we see that larger $\abs{x}$ yield smaller light ray fluctuations.

\subsubsection*{Multiple non-vanishing parameters}

\begin{figure}
    \centering
    \includegraphics[width=1.0\textwidth]{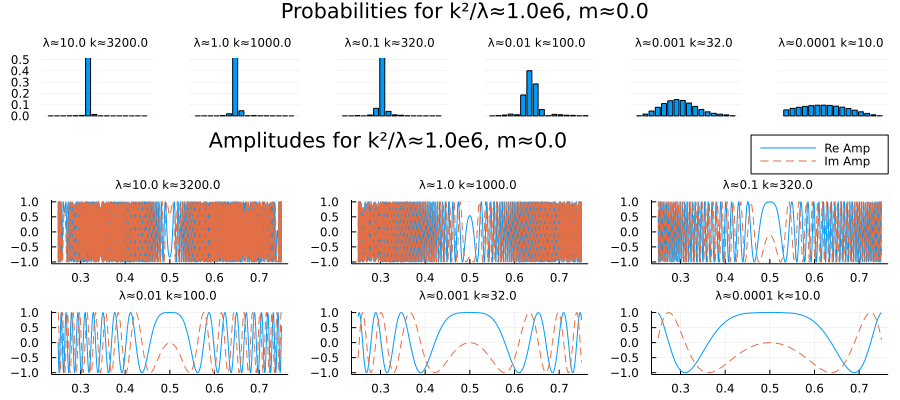}
    \caption{Probability and amplitude distributions for a family of $(\lambda,k)$ with $k^2/\lambda =10^6, m=0$.}
    \label{fig:4D_probabilities-amplitudes_lambdaa1e6}
\end{figure}

\begin{figure}
    \centering
    \includegraphics[width=1.0\textwidth]{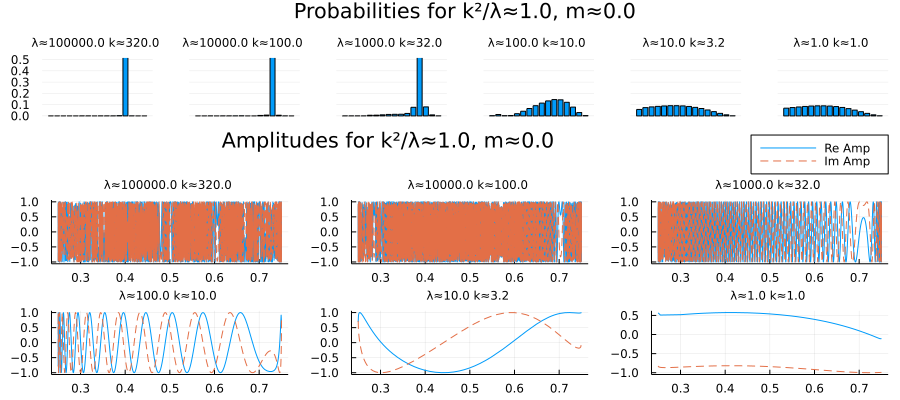}
    \caption{Probability and amplitude distributions for a family of $(\lambda,k)$ with $k^2/\lambda =1, m=0$.}
    \label{fig:4D_probabilities-amplitudes_lambdaa1}
\end{figure}

\begin{figure}
    \centering
    \includegraphics[width=1.0\textwidth]{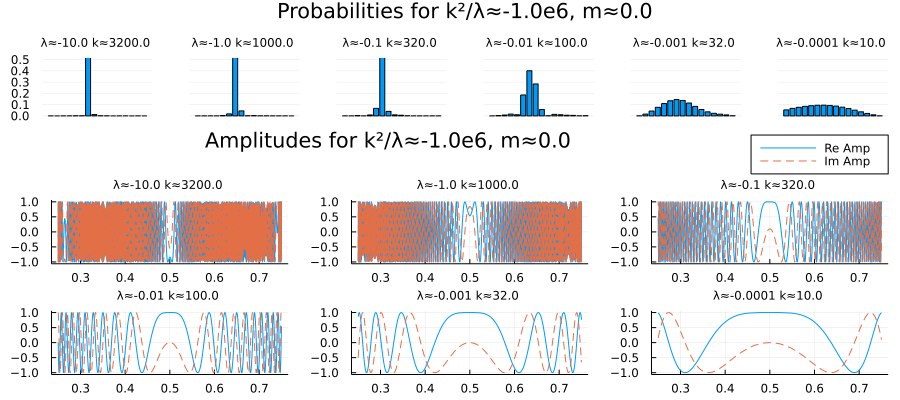}
    \caption{Probability and amplitude distributions for a family of $(\lambda,k)$ with $k^2/\lambda =-10^6, m=0$.}
    \label{fig:4D_probabilities-amplitudes_lambdaa-1e6}
\end{figure}

\begin{figure}
    \centering
    \includegraphics[width=1.0\textwidth]{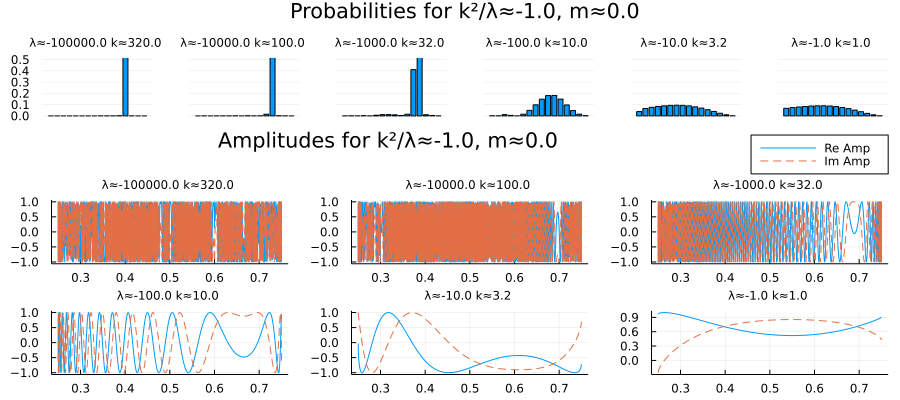}
    \caption{Probability and amplitude distributions for a family of $(\lambda,k)$ with $k^2/\lambda =-1, m=0$.}
    \label{fig:4D_probabilities-amplitudes_lambdaa-1}
\end{figure}

The results for some families of non-vanishing $(\lambda,k)$ are shown in \Cref{fig:4D_probabilities-amplitudes_lambdaa1e6} to \Cref{fig:4D_probabilities-amplitudes_lambdaa-1}. The parameters are organized according to the value of $k^2/\lambda$, which is fixed in the scaling analysis in 4D (\Cref{eq:4dvbs}). We consider $m=0$, the value for the Dewitt measure \cite{Hamber2009QuantumApproach}, when smaller values only serve to push the probability distribution towards the side like in \Cref{fig:4D_probabilities-amplitudes_m}. As in 3D light ray fluctuation is large for large values of $\lambda$ and $k$, and small for small values of $\lambda$ and $k$.

\subsection{Fixed coupling constants, varying boundary sizes}\label{eq:4dvbs}

In 4D the scaling identity \eqref{eq:si2} implies that
\begin{align}
Z[l^2\sigma_B,\lambda,k,m] =&l^{2N_e+4mN_t} ~ Z[\sigma_B,l^4\lambda,l^2k,m],\\
p_i[l^2\sigma_B,\lambda,k,a,m] =& p_i[\sigma_B,l^4\lambda,l^2 k,m].
\end{align}
Therefore all the previous figures for fixed $k^2/\lambda$ can then be read from left to right as a progressive shrinking of the size of the boundary for fixed parameters. As in 3D, light ray fluctuation increases as the boundary shrinks.

\section{Light amplitudes and the continuum limit}\label{eq:lacl}


The previous results are approximations based on a coarse simplicial lattice. To improve the approximation the lattice needs to be refined as the exact result is defined in the lattice refinement limit of \eqref{eq:pf}. However, our current understanding on how to take the continuum limit in Lorentzian quantum gravity is incomplete (see \cite{Ambjorn2020RenormalizationGeometryb} and \cite{Steinhaus2020CoarseReviewb} for related discussions on causal dynamical triangulation and spin-foams, respectively).

Therefore for further studies of light ray fluctuations we are faced with two tasks. The first is to understand light ray propagation in quantum spacetime with improved quantitative accuracy. The second is to understand the continuum limit of non-perturbative Lorentzian quantum gravity. It would be nice if the two tasks could be tackled together. In this section we discuss some thoughts along this line. The main idea is to treat light ray probabilities as the physical quantity to compare across different lattices in performing renormalization group type analysis.

\subsection{Light amplitudes}

Consider a bounded region of quantum spacetime crossed by a test light ray. Denote by $b$ the gravitational boundary condition for the region, and by $r$ the locations and directions of the light ray when it crosses the boundary. Note that a test light ray usually crosses the boundary more than once, so $r$ contains a list of variables. The \textbf{light amplitude}
\begin{align}\label{eq:la}
A[b,r]=\int_{b,r} \mathcal{D}g ~A[g]
\end{align}
is defined as the path integral over all gravitational configurations $g$ compatible with the boundary conditions $r$ and $b$.


From the light amplitudes one could derive light ray probabilities by taking the modulus square of the amplitudes, possibly after some coarse-grainings of the light ray variables $r$. This procedure is exemplified on a simple box lattice $\Gamma$ in previous sections, where we computed the coarse-grained amplitudes
\begin{align}
A_{\Gamma}[b,r_{in},i]=\int_{r_{out}\in I_i} dr ~ A_{\Gamma}[b,r]
\end{align}
for different outgoing light location intervals $I_i$ under fixed incoming light location $r_{in}$, took the modulus square, and obtained the light ray probabilities.

\subsection{Renormalization group}

The next task is to push the study to finer lattices. In lattice field theories this is usually tackled in a renormalization group analysis.

As discussed in Section 1.7 of Montvay and M{\"u}nster \cite{Montvay1994QuantumLattice}, there are two commonly adopted alternatives in lattice refinement for quantum lattice field theories. One could either fix the bare coupling constants and consider how the physical quantities such as the renormalized mass change as the lattice is refined, or fix the physical quantities and consider how the bare coupling constants change as the lattice is refined. In either case, the goal is to identify fixed points as special places in the space of the couplings where both the bare coupling constants and the physical quantities remain unchanged as the lattice is refined. In principle, the theory space comes with infinitely many coupling constants, but in practice, one often works on a subspace with a finite number of couplings. For theories exhibiting asymptotic safety, fixed points that capture the essential aspects of the full theory can be identified in the finite-dimensional subspace.

For example in scalar field theory, one works in a two-dimensional theory space of the bare mass and the bare quartic couplings \cite{Montvay1994QuantumLattice}. The renormalized mass and the renormalized quartic coupling relate more directly to laboratory observations and are picked as the physical quantities to compare across different lattices. Lattice refinement is carried out by decreasing the lattice spacing. In 4D one identifies the free theory fixed point but no interacting fixed points, which could be read as an indication that before reaching the zero lattice spacing limit some new ingredients such as additional degrees of freedom or a fundamental cutoff must come in to rescue the existence of an interacting quantum scalar field theory
\cite{Callaway1988TrivialityExist}.

For simplicial quantum gravity one could attempt a similar study. Start with an ansatz for a finite-dimensional theory space. Refine the lattice by enlarging the lattice graph, which at the expectation value level decreases the simplicial volumes $\ev{\abs{V_s}}$ when of the total spacetime volume $\ev{\sum_s \abs{V_s}}$ is bounded.

The important question is what to pick as the physical quantities to compare across lattices. Lorentzian quantum gravity reveals light ray probability as a candidate. In previous sections we computed the coarse-grained light amplitudes and probabilities
\begin{align}
A_{\Gamma,\alpha}[b,r\in I_i], \quad p_{\Gamma,\alpha}[b,r\in I_i]
\end{align}
for different bare couplings $\alpha$ on a simple lattice $\Gamma$ under symmetry-reduction. To proceed, one should try to identify fixed points in the $\alpha$ theory space where $p_{\Gamma,\alpha}$ remains unchanged as the lattice $\Gamma$ is refined. If successful, the constant physics trajectories approaching the fixed points in the theory space indicate how the continuum limit can be taken.

In this program it is crucial to find an efficient method of computation. In addition to numerical methods such as the one in \cite{Jia2022ComplexProspects}, analytical insights including ideas on how to simplify the theory without sacrificing key aspects could be helpful. Note also that the light amplitude is composable in the sense that the light amplitude on the union of multiple quantum spacetime regions can be derived from summing the products of light amplitudes for the individual regions. Combined with the scaling identities of \Cref{sec:si}, it may be possible to find some shortcuts in computing light amplitudes on refined lattices by iteratively composing light ray amplitudes on coarser lattices.

\subsection{Elementary light ray fluctuations?}

In considering the strict infinite lattice limit a question arises on light ray fluctuations in the elementary simplicies. Should we regard an elementary simplex as a flat region of spacetime where light ray fluctuations are absent, or should we regard it as representing a family of spacetime configurations in superposition where light ray fluctuations are present?

In order to satisfy quantum uncertainty relations path integrals are dominated by non-differentiable configurations (see Section 7.3 of Feynman and Hibbs \cite{Feynman1965QuantumIntegrals}).
For example, take the path integral of a non-relativistic quantum particle and suppose to the contrary that the paths are all differentiable. Then for a fixed path at a time $t$ the position $x(t)$ and momentum $p(t)$ are well-defined real numbers so that $x(t)p(t)-p(t)x(t)=0$ for this path. Since this holds for all paths,
\begin{align}
\ev{x(t)p(t)-p(t)x(t)}=0
\end{align}
as computed from the path integral. Quantum uncertainty relations would be violated, and one concludes that the assumption that the paths are differentiable cannot hold.
On the other hand, we standardly enumerate the paths in terms of piecewise linear paths \cite{Feynman1965QuantumIntegrals}. Along the interior of each linear piece of path, both the position and the momentum are well-defined real numbers. Is there not a contradiction?

\begin{figure}
    \centering
    \includegraphics[width=.5\textwidth]{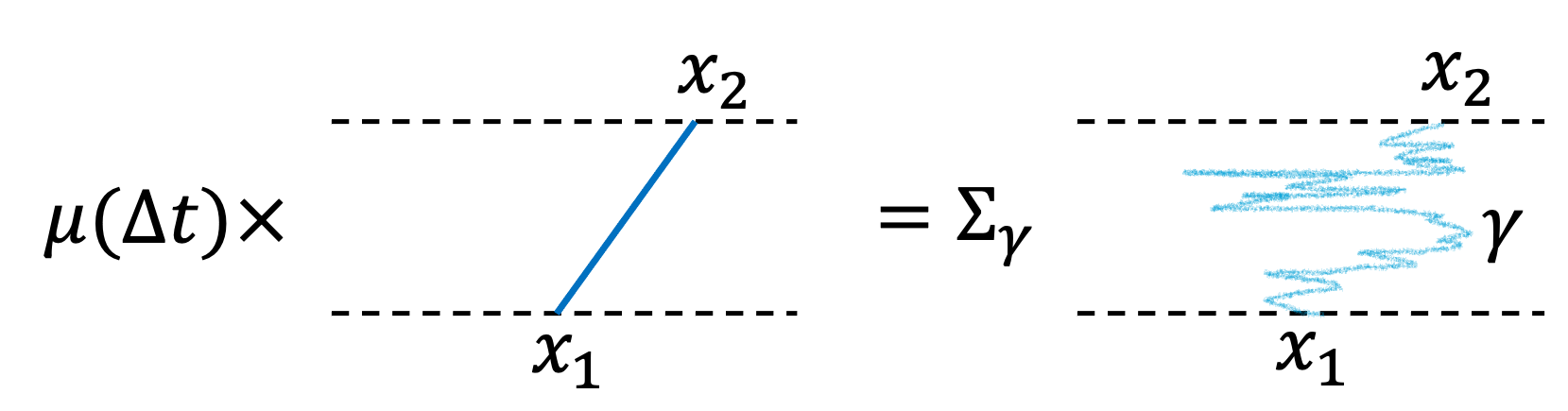}
    \caption{The path integral measure factor $\mu$ relates one linear path segment amplitude to the sum of amplitudes over a family of paths.}
    \label{fig:pi_measure}
\end{figure}

One resolution is to interpret a piecewise linear path as representing a family of paths including non-differentiable ones (\Cref{fig:pi_measure}). In the in the continuum (infinitely many time steps) limit the path integral measure factor serves to correct the difference between the amplitude of a piecewise linear path and the amplitude of family of paths including non-differentiable ones. Namely as the time interval between two steps $\Delta t\rightarrow 0$,
\begin{align}\label{eq:mfm}
\mu(\Delta t) A[\gamma_0] = \sum_{\gamma} A[\gamma]
\end{align}
as illustrated in \Cref{fig:pi_measure} for one linear piece of path.
Here $A[\gamma_0]$ is the amplitude for the linear path $\gamma_0$ connecting the starting and ending points, $\mu$ is the measure factor, and the RHS sums over paths $\gamma$ including non-differentiable ones.

For example for a free particle with mass $m$, the transition amplitude from $X_1$ to $X_2$ in time $T$ is
\begin{align}\label{eq:piaa}
\sqrt{\frac{m}{2\pi i T}} e^{im(X_2-X_1)^2/2T},
\end{align}
which can be obtained for instance from solving the Schr{\"o}dinger equation. Had we approximated the transition amplitude by the amplitude of the linear path from $X_1$ to $X_2$, it would be $e^{im(X_2-X_1)^2/2T}$, which equals the second factor of \eqref{eq:piaa}. This approximation misses all the other paths connecting $X_1$ to $X_2$, including the non-differentiable ones. Therefore the first factor of \eqref{eq:piaa} can be understood as correcting the differences in the time interval $T$. Now if we evaluate the same transition amplitude over time $T$ using the standard path integral prescription over piecewise linear paths, during a time interval $\Delta t$ a linear piece of path has amplitude $A[\gamma_0]=e^{im(x_2-x_1)^2/2\Delta t}$ between some positions $x_1$ and $x_2$. This would eventually lead to the wrong result in the $\Delta t\rightarrow 0$ limit because we missed all the other paths connecting $x_1$ to $x_2$, including the non-differentiable ones. Since \eqref{eq:piaa} applies to all time durations, correcting the difference requires the multiplication of $\sqrt{\frac{m}{2\pi i \Delta t}}$, equals the standard path integral measure factor $\mu(\Delta t)$. Hence the correct result is obtained by interpreting a piecewise linear path as representing a superposition over a family of paths whose amplitude sum differs from amplitude of the piecewise linear path by the measure factors.

A similar story can be told about simplicial gravitational path integrals which generalize the one-dimensional piecewise linear paths to higher dimensional piecewise flat simplicies. In analogy with the particle case, an elementary simplex of a simplicial spacetime configuration multiplied by the measure factor can be thought of as representing the sum over a family of configurations dominated by non-differentiable ones.

Therefore in computing light amplitudes on a finite lattice, it is reasonable to assume the presence of ``elementary light ray fluctuations''. Namely, even on an elementary simplex the light ray locations should not be sharply peaked as on a piece of Minkowski spacetime, but should exhibit quantum fluctuations due to the sum over a family of configurations. This means the light amplitudes for the elementary simplices should be non-vanishing for a range of light ray locations. 

On the other hand, it is not clear to what extent this point is practically relevant in studying the continuum limit. It could be that once the lattice is taken large enough so that the elementary simplicies are taken small enough (in the sense of having small expectation value for the volume), introducing elementary light ray fluctuations or not does not affect the results. Further studies are needed to see if this is the case.

\section{Discussion}\label{sec:d}

In this work we studied light ray fluctuations in a simple model in Lorentzian simplicial quantum gravity. The overall question is how a quantum region of spacetime affects the propagation of light rays crossing it. We computed the probabilities for test light rays to land at different locations across a symmetry-reduced box model with simple boundary conditions in 2,3 and 4 spacetime dimensions.

For fixed boundary conditions light ray fluctuations are generically large when all coupling constants are relatively small in absolute value in all dimensions. In contrast, for fixed coupling constants light ray fluctuations show different trends in different dimensions when the boundary size is decreased. While in 2D light ray fluctuations first increase and then decrease, in 3D and 4D light ray fluctuations just increase without decreasing. The difference can be understood by noting that the coupling constants of the cosmological constant, Einstein-Hilbert, and $R^2$-terms have different length dimensions in different spacetime dimensions, so behave differently as the boundary length is scaled as in \Cref{sec:si}.

The symmetry-reduced box models with simple boundary conditions can be generalized in two directions. Firstly we could allow more dynamical degrees of freedom by relaxing the symmetry-reduction assumption and/or introducing larger simplicial lattices. This generalization is a necessity if we are to investigate the continuum limit of the theory as discussed in \Cref{eq:lacl} where we propose to explore the use light amplitudes in renormalization group type analysis for Lorentzian quantum gravity. The continuum limit in turn would allow us to fix the coupling constants by comparing with empirical data (see \cite{Hamber2019VacuumGravity} for a related discussion in Euclidean simplicial quantum gravity) in order to turn the qualitative conclusions about the amount of light ray fluctuation into quantitative predictions. Secondly we could consider additional boundary conditions. Of particular interest are tunneling boundary conditions that admit complex, non-Lorentzian stationary points which are of interest to singularity resolution (see \cite{Jia2022ComplexProspects} and references therein). Because the path integrals are not dominated by particular Lorentzian configurations, light ray fluctuations could be generically large for singularity resolving tunneling processes.

\section*{Acknowledgement}

I am very grateful to Lucien Hardy and Achim Kempf for long-term encouragement and support. Research at Perimeter Institute is supported in part by the Government of Canada through the Department of Innovation, Science and Economic Development Canada and by the Province of Ontario through the Ministry of Economic Development, Job Creation and Trade.

\bibliographystyle{unsrt}
\bibliography{mendeley.bib}

\end{document}